\documentclass[12pt,preprint]{aastex}

\usepackage{amsmath}

\received{2005 February 15}
\begin{document}

\title{SNSPH: A Parallel 3-D Smoothed Particle Radiation Hydrodynamics  
  Code}  
  
\author{Christopher L. Fryer$^{1,2}$, Gabriel Rockefeller$^{1,2}$, and  
  Michael S.  Warren$^{1}$}  
  
\altaffiltext{1}{Theoretical Astrophysics, Los Alamos National  
  Laboratory, Los Alamos, NM 87545}  
  
\altaffiltext{2}{Department of Physics, The University of Arizona,  
  Tucson, AZ 85721}  
  
\begin{abstract}   
  We provide a description of the SNSPH code---a parallel  
  3-dimensional radiation hydrodynamics code implementing treecode  
  gravity, smooth particle hydrodynamics, and flux-limited diffusion  
  transport schemes.  We provide descriptions of the physics and  
  parallelization techniques for this code.  We present performance  
  results on a suite of code tests (both standard and new), showing  
  the versatility of such a code, but focusing on what we believe are   
  important aspects of modeling core-collapse supernovae.  
  
\end{abstract}  
  
  
\section{Introduction}  
  
SNSPH is a particle-based code, using smooth particle hydrodynamics  
(SPH) to model the Euler equations and a flux-limited diffusion  
package to model radiation transport.  Its tree algorithm is designed  
for fast traversal on parallel systems (Warren \& Salmon 1993,1995)  
and has been shown to scale well up to high processor number on a wide  
variety of computer architectures.  The increased synchronization  
required for the transport algorithm limits this scalability to  
roughly 512 processors on current parallel computers for the 1-10  
million particle core-collapse calculations that are feasible with our  
current codes and current computing power.  
  
SNSPH has now been used in several papers appearing in the literature  
studying a range of problems from stellar collapse (Fryer \& Warren  
2002,2004; Fryer 2004; Fryer \& Kusenko 2005) to supernova explosions  
and supernova remnants (Hungerford et al. 2003, 2005; Young et  
al. 2005; Rockefeller et al. 2005a; Fryer et al. 2005) to models of  
massive binaries and winds from these binaries (Fryer \& Heger 2005;  
Fryer, Rockefeller, \& Young 2005) to models of both the gas and disk  
evolution in the Galactic Center (Rockefeller et al. 2004, 2005b,  
2005c).  But none of these papers has provided a detailed description  
of the SNSPH code itself.  Here we describe these details: \S 2  
describes much of the physics implementation from gravity to  
hydrodynamics to radiation transport; and \S 3 describes many of the  
numerical techniques used in this particle-based code, focusing on  
many of the computational techniques required to make such a scheme  
scalable.  In \S 4, we show the results of SNSPH for a number of tests  
(both standard and new) used to confirm the validity of SNSPH for  
core-collapse simulations.  Astrophysics problems tend to be too  
complex for most standard code tests to completely confirm the  
validity of the code for that specific astrophysics problem.  We  
present a small suite of tests to show the broad range of physics that  
must be tested to model core-collapse.  These tests demonstrate the  
wide applicability of this SNSPH code, but further testing is required  
to truly test all the physics in the code.  We conclude with a list of  
strengths and weaknesses of SNSPH.  
  
\section{Physics Implementation}  
  
As with many astrophysics problems, solving the supernova problem  
requires a wide range of physics.  This physics must be, in one way or  
another, implemented into the numerical code simulating this  
phenomena.  In this section, we describe the major aspects of the  
physics incorporated into SNSPH: gravity, hydrodynamics (including the  
equation of state), and radiation transport (including opacities).    
Where possible, we have sought to make the code modular to facilitate   
modifications in the different microphysics packages from the equation   
of state to neutrino cross-sections and emission routines.  
  
\subsection{Gravity}  
  
Newton's second law of motion and law of gravitation provide an  
expression for the acceleration of one body under the combined  
gravitational influence of a set of $N-1$ other bodies, according to  
\begin{equation}  
\frac{d^2 \vec{x}_i}{{dt}^2} = \sum_{j \ne i}^N \vec{a}_{ij} = \sum_{j \ne i}^N \frac{G m_j \vec{r}_{ij}}{{|\vec{r}_{ij}|}^3},  
\end{equation}  
where $\vec{r}_{ij} \equiv \vec{x}_i - \vec{x}_j$.  Using this  
equation to calculate the acceleration for one body requires $N - 1$  
evaluations of the term inside the sum, so determining the  
accelerations of all $N$ bodies in a simulation requires $N (N - 1)$  
or $O(N^2)$ operations.  Performing this type of pairwise summation to  
calculate gravitational interactions is prohibitively expensive for  
all but the smallest sets of bodies, even on the fastest  
supercomputers.  
  
A number of approximate methods have been developed to calculate  
gravitational forces among large numbers of bodies by considering as  
one interaction the total effect of a set of bodies on one body, or  
the effect of one set of bodies on another set, with time requirements  
that scale as $O(N \log N)$ or $O(N)$, respectively.  Codes that use  
adaptive tree structures to subdivide the volume of a simulation and  
distinguish between nearby and distant bodies can easily implement  
such accelerated techniques; our code uses such a tree and is  
therefore one of a class of codes called ``treecodes''.  This class  
includes many SPH codes such as Gadget \cite{SpringelV:GADcca} and  
Gasoline \cite{WadsleyJW:Gasfpi} as well as particle mesh codes  
\cite{Sur04} and some Adaptive Mesh Refinement Codes such as RAGE  
\cite{Cok05}.  
  
Knowledge of the spatial arrangement of bodies in a simulation allows  
the treecode to distinguish between ``nearby'' bodies, for which  
direct pairwise calculation of gravitational forces is appropriate,  
and ``distant'' bodies, for which an approximate technique will yield  
a sufficiently accurate value for the force.  The simplest  
approximation combines a set of distant bodies into one object with a  
total mass equal to the sum of the masses of the individual bodies in  
the group, positioned at the location of the center of mass of the  
group:  
\begin{equation}\label{eqn:multipole}  
\sum_{j \ne i} - \frac{G m_j \vec{r}_{ij}}{{|\vec{r}_{ij}|}^3} \approx - \frac{G M \vec{r}_{i,cm}}{{|\vec{r}_{i,cm}|}^3},  
\end{equation}  
where $\vec{r}_{i,cm} = \vec{r}_i - \vec{r}_{cm}$, $\vec{r}_{cm}$ is  
the location of the center of mass of the group of bodies, and $M$ is  
the total mass of the group.  In principle this equation could include  
additional terms on the right-hand side to account for the quadrupole  
and higher moments of the mass distribution.  In practice, equivalent  
accuracy and higher performance is obtained for moderate levels of  
accuracy ($\sim$ 0.1 percent) by including only the  
monopole contribution (and, implicitly, the dipole contribution, which  
is equal to zero) and using an appropriate criterion for determining  
when the approximation is accurate enough.  
  
Some criterion must be used to determined when the multipole  
approximation in equation~\ref{eqn:multipole} is sufficiently accurate  
to use instead of direct summation of the accelerations due to each  
body in the set.  This criterion is called the ``multipole  
acceptability criterion'' or MAC.  Many different MACs have been  
proposed, and several are in widespread use.  Salmon \& Warren (1994)  
analyzed the worst-case behavior of several different MACs and  
developed a technique to determine a strict upper limit on the errors  
in each acceleration calculation.  As described in (Warren \& Salmon  
1993,1995), we use a MAC that incorporates such an error estimate into  
the calculation of a critical radius, according to  
\begin{equation}  
r_{c} \geq \frac{b_{max}}{2} + \sqrt{ \frac{{b_{max}}^2}{4} + \sqrt{ \frac{3 B_2}{\Delta_{interaction}} } },  
\end{equation}  
where $b_{max}$ is the size of the cell and $B_2 = \sum_i m_i  
{|\vec{r}_i - \vec{r}_{cm}|}^2$ is the trace of the quadrupole moment  
tensor.  A body and cell separated by a distance greater than $r_c$  
can use equation~\ref{eqn:multipole} to evaluate the acceleration, and  
the absolute error is guaranteed to be less than  
$\Delta_{interaction}$; a body and cell closer than $r_c$ must use  
pairwise summation for each body in the cell, or subdivide the cell  
into smaller cells and reconsider the interactions, to ensure that the  
error is not larger than that tolerance.  
  
\subsection{SPH Hydrodynamics}  
\label{Code:Hydro}  
  
The particle-based structure of our code allows us to easily implement  
the smooth particle hydrodynamics (SPH) to model the Euler (inviscid)  
equations.  SPH, invented in 1977 (Lucy 1977, Gingold \& Monaghan  
1977), has become the primary multi-dimensional Lagrangian technique  
used in astrophysics.  Its versatility allows it to be used on a  
variety of astrophysics problems (see Benz 1988; Monaghan 1992 for a  
review).  Many variants of SPH have been developed and a number of  
excellent reviews on the SPH technique, and its variations, already  
exist (Benz 1989; Monaghan 1992; Morris 1996; Rasio 1999; Monaghan  
2005); we provide a brief review here and direct readers interested in  
more details to the above reviews.  Our code was developed using the  
Benz version of SPH (Benz 1984, 1988, 1989) as a model and is nearly  
identical to those codes based on this version of SPH.  
  
\subsubsection{Brief SPH Primer}  
  
SPH is a particle based method where particles act as interpolation  
points to determine matter conditions throughout the simulation space.  
Consider the following integral representation of the quantity $A$:  
\begin{equation}  
A_s(\vec{r})=\int A(\vec{r}') W(\vec{r}-\vec{r}',h) dr'  
\end{equation}  
where $W(\vec{r}-\vec{r}',h)$ (the ``kernel'') has the following properties:  
\begin{equation}  
\int W(\vec{r}-\vec{r}',h)dr' = 1  
\label{eq:kernelint}  
\end{equation}  
and  
\begin{equation}  
\lim_{h \to 0} W(\vec{r}-\vec{r}',h) = \delta(\vec{r}-\vec{r}').  
\label{eq:kernellim}  
\end{equation}  
By integrating $A$ with our kernel, $A_s$ is the ``smoothed'' version  
of $A$ (hence the origin of the ``Smooth'' in SPH).  Note that $A_s$  
approaches $A$ as $h \to 0$.  If we expand $A(\vec{r})$ in a Taylor  
series, we find that:  
\begin{equation}  
 A_s(r)=A(r)+c(\nabla^2 A)h^2 + O (h^3)  
\end{equation}  
This SPH formulism introduces an error of order $h^2$ in the estimate of  
the quantity $A(r)$.  Discretizing this method, the integral over $A$  
is reduced to a summation over a number of points (particles) in  
space:  
\begin{equation}  
A_s(r)=\sum_j A_j (m_j/\rho_j) W(\vec{r}-\vec{r}_j,h)  
\end{equation}  
where $A_j$, $m_j$, and $\rho_j$ are the respective values of $A$, the  
mass, and the density of particle $j$.  The structure of the smoothing  
is determined by the kernel $W(\vec{r}-\vec{r}',h)$ where $h$ denotes  
the size over which the smoothing occurs (see below).  
  
Although any kernel will work as long as it satisfies equations  
\ref{eq:kernelint} and \ref{eq:kernellim}, determining the best kernel  
for a given problem can be a black art.  One of the simplest kernels, and the   
one we use for most of our calculations (although we also use other spline   
kernels) is a cubic spline kernel:  
\begin{equation*}  
W(r,h)=\pi^{-1} h^{-3}  
\begin{cases}   
 1-1.5v^2+0.75v^3 & \text{if $0 \leq v \leq 1$,} \cr  
 0.25(2-v)^3  & \text{if $1 \leq v \leq 2$,} \cr  
 0. & \text{otherwise,}   
\end{cases}  
\end{equation*}  
where $v=r/h$.  With this kernel, a given interpolation point  
(particle) contributes to the value of $A(r)$ only if $r$ is within  
$2h$ of that particle.  The value of $h$ for a given particle $i$,  
termed the ``smoothing length'', is allowed to vary with time using  
the relation presented in Benz (1989):  
\begin{equation}  
dh_i/dt = -1/3 (h_i/\rho_i) (d\rho_i/dt)  
\end{equation}  
This variation is necessary to ensure full spatial coverage by the  
particles (we would like any position to overlap with a base number of  
particles) and, as long as $h$ varies on a scale similar to other  
variables, the errors remain of $O(h^2)$ (Hernquist \& Katz 1989).  We  
define the interaction between a particle $i$ and a neighboring  
particle $j$ by evaluating a mean value $h \equiv (h_i+h_j)/2$.  We  
additionally set limits for the number of neighbors (the standard  
range for our 3-dimensional models is between $\sim$40 and $\sim$80  
neighbors).  In the extreme case that the number of neighbors falls  
above (or below) these maximum (minimum) values, we additionally lower  
(raise) $h_i$ by a configurable amount on each timestep (typically a  
factor of 0.002-0.1) to enforce this range of neighbors (this occurs  
rarely, if at all, for a given particle during the course of a  
simulation).  
  
We can now use our interpolation points to calculate the value of any  
quantity and its derivatives over our spatial domain.  The   
density at position $i$ is simply:  
\begin{equation}  
\rho_i= \sum_j m_j W(\vec{r}_i-\vec{r}_j,h).  
\end{equation}  
As with any numerical technique, there is more than one way to  
discretize our system.  This is most apparent in the calculation of  
derivatives.  For example, in principal, the gradient of a function  
$A$ at particle $i$ is just:  
\begin{equation}  
\nabla A_i = \sum_b A_j (m_j/\rho_j) \nabla W(\vec{r}_i-\vec{r}_j,h).  
\end{equation}  
In practice, it is more accurate to use,  
\begin{eqnarray}  
\label{eq:trick}  
\nabla A_i & = & 1/\rho [\nabla(\rho A) - A \nabla\rho] \\ & = &  
          1/\rho_i \sum_j (A_j - A_i) m_j \nabla  
          W(\vec{r}_i-\vec{r}_j,h).  
\end{eqnarray}  
The Morris (1996) review includes quite a bit of discussion about   
these ``techniques'' used to improve SPH, also noting the situations   
when one technique might be better than another.  We follow the Benz   
version of SPH for all our discretization assumptions in solving   
the hydrodynamics equations.   
  
\subsubsection{Continuity Equation}  
  
The Benz version of SPH is a true Lagrangian code - the mass and  
number of particles is conserved, so the total mass in the system is  
also conserved.    
  
\subsubsection{Momentum Equation}  
  
We evaluate momentum and energy conservation for the particles   
themselves and assume an inviscid gas.  Hence, the hydrodynamic   
equations reduce to the Lagrangian form of the Euler equations:  
\begin{equation}  
d\vec{v}_i/dt=-1/\rho_i (\nabla P)_i  
\end{equation}  
where $\vec{v}_i=d\vec{r}_i/dt$ and $P_i$ is the pressure of particle  
$i$.  If we simply use equation \ref{eq:trick}, the pressure gradient  
for particle $i$ is written:  
\begin{equation}  
(\nabla P)_i = 1/\rho_i \sum_j m_j (P_j - P_i) \nabla_i W_{ij}  
\end{equation}  
where $W_{ij} = W(\vec{r}_i-\vec{r}_j,h)$.  Although such a scheme can  
be used, it is not symmetric (and hence does not conserve linear and  
angular momentum).  We can instead write:  
\begin{equation}  
\nabla P/\rho = \nabla (P/\rho) + P/\rho^2 \nabla \rho  
\end{equation}  
Now using equation \ref{eq:trick} on this representation, we   
get:  
\begin{equation}  
\label{eq:mom}  
d\vec{v}_i/dt = - \sum_j m_j (P_i/\rho_i^2 + P_j/\rho_j^2) \nabla_i W_{ij}.  
\end{equation}  
Let's confirm that this algorithm conserves linear momentum.  The  
force on particle $i$ due to particle $j$ is equal to the negative  
force on particle $j$ due to particle $i$:  
\begin{equation}  
\label{eq:momcons}  
m_i d\vec{v}_i/dt = m_i m_j (P_i/\rho_i^2 + P_j/\rho_j^2) \nabla_i W_{ij}  
= - m_j d\vec{v}_j/dt,  
\end{equation}  
where we have taken advantage of the fact that the kernels are  
anti-symmetric $\overrightarrow{\nabla}_i  
W_i(|\overrightarrow{r}_i-\overrightarrow{r}_j)|,h) = -  
\overrightarrow{\nabla}_j  
W_j(|\overrightarrow{r}_i-\overrightarrow{r}_j)|,h)$.  We will show  
that angular momentum is conserved in section \ref{sec:angm}.  
  
\subsubsection{Energy Conservation}  
  
Most hydrodynamics codes evolve either the total energy (kinetic +  
internal) or the internal (thermal) energy alone.  In a system with  
gravity, the total energy also can include the gravitational potential  
energy.  Evolving the total energy ensures a better conservation of  
the total energy.  But in astrophysics, accurate temperatures (and  
hence energies) are important even in cases where the gravitational  
potential and kinetic energies dominate the total energy by many  
orders of magnitude.  To obtain reliable internal energies, it is  
often better to evolve the internal energy alone in the energy  
equation.  The evolution of the specific internal energy $u$ is  
\begin{equation}  
du/dt = -(P/\rho) \nabla \cdot \vec{v},  
\end{equation}  
corresponding to the following SPH formulation:  
\begin{equation}  
du_i/dt = P_i/\rho_i^2 \sum_j m_j (\vec{v}_i-\vec{v}_j) \cdot \nabla_i W_{ij}  
\end{equation}  
See Morris (1996) for other valid SPH formulations of the energy  
equation.  
  
It is easy to show that this formulation combined with our momentum   
equation conserves total energy (kinetic and thermal energies).  The   
total energy for all particles is given by:  
\begin{equation}  
d/dt \sum_i m_i u_i = \sum_i \sum_j m_i m_j P_i/\rho_i^2 (\vec{v}_i-\vec{v}_j)   
\cdot \nabla_i W_{ij}  
\end{equation}  
By interchanging indices and again making use of the identity $\nabla_i W_{ij}   
= - \nabla_j W_{ij}$, we get  
\begin{equation}  
d/dt \sum_i m_i u_i = \sum_i \sum_j m_i m_j (P_i/\rho_i^2 +  
P_j/\rho_j^2) (\vec{v}_i-\vec{v}_j) \cdot \nabla_i W_{ij}.  
\end{equation}  
Comparing this to equation \ref{eq:mom}, we find:  
\begin{equation}  
d/dt \sum_i m_i u_i = d/dt (1/2 \sum_i m_i v_i^2)  
\end{equation}  
which shows that the work done by pressure forces changing the kinetic  
energy comes at the expense of the internal energy, ensuring the  
conservation of total energy.  
  
In core-collapse, it is often better to follow the entropy, instead of  
the internal energy, of matter.  For degenerate matter, the  
temperature can vary wildly over small changes of the internal energy.  
Entropy varies more rapidly with temperature in degenerate conditions.  
By using entropy as the energy parameter, we get more stable  
temperature values.  We use the same hydrodynamics equations, setting  
$ds_i/dt = (1/T) (du_i/dt)$ where $T$ is the matter temperature.  
  
\subsubsection{Artificial Viscosity}  
  
We have limited our description of hydrodynamics to inviscid (Euler)  
equations.  It is well known that any Euler method with finite  
resolution is unable to describe shocks and will result in large,  
unphysical, oscillations unless one includes some sort of viscosity,  
low order diffusion or a Riemann solver.  Although Godunov-type  
methods have been developed for SPH (Inutsuka 1994; Monaghan 1997;  
Inutsuka 2002), most techniques introduce a viscosity term to handle  
shocks (Benz 1989; Monaghan 1992, Monaghan 2005):  
\begin{equation*}  
\Pi_{ij}=  
\begin{cases} (-\alpha \bar{c} \mu_{ij} + \beta \mu_{ij}^2)/\bar{\rho}_{ij}   
& \text{if $(\vec{v}_i-\vec{v}_j) \cdot (\vec{r}_i-\vec{r}_j) \leq  
  0$,} \cr   
0 & \text{otherwise,}   
\end{cases}  
\end{equation*}  
where  
\begin{equation}  
\mu_{ij} = \frac{h (\vec{v}_i-\vec{v}_j) \cdot (\vec{r}_i-\vec{r}_j)}  
{|\vec{r}_i-\vec{r}_j|^2 + \epsilon h^2},  
\end{equation}  
$\bar{\rho}_{ij} = 1/2 (\rho_i+\rho_j)$ and $\bar{c}_{ij} = 1/2  
(c_i+c_j)$ are the average of the densities and sound speeds of the  
interacting particles, $\alpha$ and $\beta$ are the bulk and von  
Neumann-Richtmyer viscosity coefficients respectively (typically set  
to 1.5 and 3.0), and $\epsilon$ is a factor to avoid   
divergences at small separations (typically set to 0.01).  
  
The momentum and energy (internal + kinetic) equations can now be  
rewritten to include this viscosity:  
\begin{equation}  
d\vec{v}_i/dt = - \sum_j m_j (P_i/\rho_i^2 + P_j/\rho_j^2 + \Pi_{ij})  
\nabla_i W_{ij}  
\end{equation}  
and  
\begin{equation}  
du_i/dt = P_i/\rho_i^2 \sum_j m_j (\vec{v}_i-\vec{v}_j) \cdot \nabla_i W_{ij}  
+1/2 \sum_j m_j \Pi_{ij} (\vec{v}_i-\vec{v}_j) \cdot \nabla_i W_{ij}.  
\end{equation}  
Clearly, the addition of the artificial viscosity term retains our   
total (kinetic + internal) energy conservation.  
  
\subsubsection{Equation of State}  
  
To complete these equations, we must include an equation of state to  
determine pressures from internal energies or entropies.  Our basic  
SPH scheme includes equations of state for isothermal and ideal gases.    
But we have also included a number of equations of state and the code   
has thusfar not encountered any problems incorporating new equations of   
state.  
  
The most complex equation of state we have in the code is the one we  
have used in most of our supernova simulations.  For the core-collapse  
problem we use an equation of state combining the nuclear equation of  
state by Lattimer \& Swesty (1991) at high densities and the Blinnikov  
et al. (1996) equation of state at low densities.  Nuclear burning is  
approximated by a nuclear statistical equilibrium scheme (Hix \&  
Thielemann 1996).  
  
The Lattimer \& Swesty (1991) equation of state can be used from  
densities below $10^9 {\rm g \, cm^{-3}}$ up to densities above  
$10^{15} {\rm g \, cm^{-3}}$.  However, an error in the energy levels  
of this equation of state cause it to give incorrect answers below  
$10^{11} {\rm g \, cm^{-3}}$ (Timmes et al.  2005) and be aware that  
we do not know the true behavior of matter above nuclear densities and  
we expect this part of the equation of state to change with time.  It  
is valid for electron fractions from 0.03 up to 0.5.  If the electron  
fraction exceeds this value, we assume the pressure is set to 0.5.  In  
typical simulations, we use the Lattimer \& Swesty equation of state  
for matter above $10^{11} {\rm g \, cm^{-3}}$.  Below $10^{11} {\rm g  
\, cm^{-3}}$, we use the Blinnikov equation of state.  This equation  
of state is valid to densities as low as $100 {\rm g \, cm^{-3}}$.  
Below these densities, the equation of state is comparable to an ideal  
gas equation of state.  
  
The nuclear statistical equilibrium scheme is used for material in the  
Blinnikov equation of state with temperatures above $5 \times 10^{9}$K  
(depending upon the problem, this value is sometimes set a factor of 2  
lower or higher).  This scheme uses 16 species, focusing on the  
iron-peak elements from $^{54}Fe$ to $^{86}$Kr with typical Q-values  
for reactions among stable nuclei heavier than silicon lying between  
8-12MeV.  The version of the equation of state we describe here is   
not tabular, but a set of analytic function calls.  Timmes et al. (2006)   
have developed a tabular version of this equation of state which is   
currently being tested.  
  
We have added small networks, such as the 14-element alpha network by  
Benz, Thielemann, \& Hills (1989) to the code to follow the burning in  
the non-steady state.  In general, the burning time-step is much  
shorter than the hydrodynamic timestep, so the burning is done in a  
sub-cycle using Newton-Raphson iterations to converge.  For most  
core-collapse calculations where we include this network, burning is  
only considered for temperatures above $8 \times 10^{7}$K and the  
network is switched over to nuclear statistical equilibrium at $5  
\times 10^{9}$K.  Comparisons with large networks have shown   
that this network does not produce accurate yields, but the   
energies are reasonable (Young et al. 2005).  
  
We have also added an equation of state to model the basalt and iron  
material in planetary cores (Tillotson 1962).  The addition of new  
equations of state is a straightforward exercise ($<1$ day of work).  
  
\subsection{Radiation Transport}\label{Code:RadTran}  
  
The radiation transport scheme currently implemented in the SPH code  
is based on the 2-dimensional explicit, flux-limited transport scheme  
developed by Herant et al. (1994).  Flux-limited diffusion is a moment  
closure technique where the equations are closed in the first moment.  
For neutrino diffusion, we transport the neutrino number and the  
advection term in the flux limited diffusion equation is simply:  
\begin{equation}  
1/c \partial n_{\nu}/\partial t = \Lambda \nabla n_{\nu},  
\end{equation}  
where $n_{\nu}$ is the neutrino number, c is the speed of the   
neutrinos (set to the speed of light) and $\Lambda$ is the   
flux-limiter.  Here we describe the 3-dimensional  
adaptation of this scheme.  We also comment on a number of  
peculiarities of this scheme that should be understood before its use.  
  
In the current version of the code, we consider the transport of 3  
neutrino species ($l=\nu_e,\bar{\nu_e},\nu_x$ where $\nu_x$  
corresponds to the $\tau, \mu$ neutrinos and their anti-particles that  
are all treated equally).  Because neutrino number is the conserved  
quantity, we transport neutrino number and then determine the energy  
transport by using the mean neutrino energies.  The radiation  
transport scheme in our SPH code is modeled after the technique to  
calculate forces in SPH: we calculate symmetric interactions  
between all neighbor particles.  Hence, our flux-limited diffusion  
scheme calculates the radiation diffusion in or out of a particle by  
summing the transport over all neighbors (the equivalent of all  
bordering cells in a grid calculation).  The neutrino transport for  
particle i is given by:  
\begin{equation}  
dn^i_{\nu_l}/dt = \sum_j \Lambda^{ij}_{\nu_l} (n^j_{\nu_l}   
b^{j \rightarrow i}_{\nu_l} - n^i_{\nu_l} b^{i \rightarrow j}_{\nu_l})  
\nabla W^{ij} m^j/\rho^j  
\end{equation}  
and the corresponding energy transport is:  
\begin{equation}  
de^i_{\nu_l}/dt = \sum_j \Lambda^{ij}_{\nu_l} (\epsilon^j_{\nu_l}  
n^j_{\nu_l} b^{j \rightarrow i}_{\nu_l} - \xi^{i \rightarrow j}  
\epsilon^i_{\nu_l} n^i_{\nu_l} b^{i \rightarrow j}_{\nu_l}) \nabla  
W^{ij} m^j/\rho^j  
\end{equation}  
where $n^i_{\nu_l}, e^i_{\nu_l}$ are, respectively, the neutrino  
density and energy density in particle i for species $\nu_l$,  
$\epsilon_{\nu_l}^i$ is the mean neutrino energy (an average energy  
taking into account the $\epsilon_{\nu_l}^2$ dependence of the   
neutrino opacity), and $\xi^{j \rightarrow i}$ is the redshift  
correction for $\epsilon_{\nu_l}^j$ as seen by particle i.    
$b^{i \rightarrow j}_{\nu_l}$ are the fermion blocking factors   
for neutrinos.  
  
$\Lambda^{ij}_{\nu_l}$ is effectively the limiter for the flux-limited  
transport scheme (we have modified the definition slightly to fit in  
our numerical equations).  The simplest such scheme for 3-dimensions  
is:  
\begin{equation}  
\Lambda^{ij}_{\nu_l} = min (c, D^{ij}_{\nu_l}/r^{ij})  
\end{equation}  
where $c$ is the speed of light,  
$D^{ij}_{\nu_l}=2D^i_{\nu_l}D^j_{\nu_l}/(D^i_{\nu_l}+D^j_{\nu_l})$ is  
the harmonic mean of the diffusion coefficients for the species  
$\nu_l$ of particles i and j and $r^{ij}$ is the distance between  
particles i and j.  This limiter was used by Herant et al. (1994) and,  
for comparison with Herant et al. (1994), by Fryer \& Warren (2002),  
but we have used a number of other flux-limiters, all of which are  
valid under this transport scheme (Fryer et al. 1999).  The opacities  
used to determine the diffusion coefficients are given in Herant et  
al. (1994).  
  
Beyond some radius in a core-collapse simulation, neutrinos are  
essentially in the free-streaming regime where transport is not  
necessary (unless one wants to truly follow the radiation wave as it  
progresses through the star).  We do not model transport beyond this  
``trapping'' radius.  Instead we sum up all neutrinos that transport  
beyond this radius and emit them using a light-bulb approximation.  
That is, the material beyond this radius sees a constant flux and we  
can determine the amount of energy a particle gains ($dE_i/dt$) from  
neutrino interactions simply by using the free-streaming limit:  
\begin{equation}  
dE_i/dt = L_\nu \left(1.0-e^{-\Delta \tau_i} \right)  
\end{equation}  
where $L_\nu$ is the neutrino luminosity and $\Delta \tau_i$ is the  
optical depth of a particle $i$.  This assumption is only valid if the  
total amount of energy imparted ($\sum_i dE_i/dt$) from the neutrinos  
onto the matter is much less than the total neutrino flux ($L_\nu$).  
To guarantee this, we determine this trapping radius by evolving it  
with time such that $(\sum_i dE_i/dt)/L_\nu$ is always less than some  
value.  This value was originally set to 0.1 by Herant et al. (1994),  
but in recent calculations, we use 0.05.  
  
Such a scheme can be easily converted into multi-group, but such  
modifications have not yet been done.  The scheme scales reasonably  
well on multiple processors (for a 5 million particle run, the code  
has scaled nearly linearly up to 256 processors on the Space Simulator  
Beowulf cluster and up to 512 processors on the ASC Q machine at Los  
Alamos National Laboratory).  In part, this scalability is due to the  
explicit nature of the transport scheme.  In general, explicit  
transport schemes strongly limits timesteps as the speed of light, not  
sound, constrains the duration of the timestep.  In core-collapse  
supernovae, this constraint is not too onerous because the sound speed  
is nearly a third the speed of light anyway, so the explicit transport  
scheme leads to only a factor of 3 decrease in the timestep.  But this  
explicit flux-limited transport scheme can be used in a much wider  
variety of problems where the mean free path is very short for the  
smallest particles.  Such scenarios occur in many astrophysics  
problems.  
  
Before we show how well SNSPH performs on our suite of test problems,   
let's describe the computational techniques used to make this code run   
efficiently on parallel architectures.  
  
\section{Computational Issues}  
  
On a single processor, careful attention to the computational details  
can significantly accelerate a code's performance.  For large-scale  
parallel architectures, these details are critical to taking full  
advantage of these supercomputers.  Unfortunately, the more complex  
the code becomes, the more ingenuous the computational techniques must  
be to preserve scalability.  Here we discuss just the basic  
computational issues, focusing on our tree algorithm (see also Warren  
\& Salmon 1993, 1995) and the basic parallelization issues arising  
from use of this tree.  We conclude with a discussion of the timestep  
integrator.  
  
\subsection{Treecode}\label{Code:Tree}  
  
The calculation of gravitational forces between bodies in a N-body  
simulation and the identification of neighbors in a SPH calculation  
are both accelerated significantly through the use of a treecode, in  
which a hierarchical ``tree'' data structure is constructed to  
represent the spatial arrangement of bodies in a simulation.  Our code  
uses a ``parallel hashed oct-tree algorithm'' (Warren \& Salmon 1993,  
1995), in which each node in the tree can have up to eight  
``daughter'' nodes below it.  The ``root'' of the tree (often imagined  
to be at the top of the tree in computer science discussions)  
represents the entire volume of a simulation; each of the root's eight  
daughter nodes represents one octant of that overall volume.  A  
complete hierarchical representation of the volume can be created by  
recursively subdividing each octant and adding more levels of daughter  
nodes to the tree until some stopping condition is reached.  In our  
implementation of a treecode, we stop subdividing a volume when that  
volume contains zero particles or only one particle; the node  
representing that volume in the tree is called a ``leaf'' node and has  
no daughters of its own.  By performing traversals of the hierarchical  
tree structure, the treecode can rapidly distinguish between nearby  
and distant particles and accelerate the calculation of gravitational  
and pressure forces in simulations.  
  
In descriptions of N-body techniques, the term ``body'' generally  
refers to one of the fundamental entities being simulated.  In  
smoothed particle hydrodynamics, the interacting entities are  
logically called ``particles''.  In the following discussion, the  
terms ``particle'' and ``body'' are used interchangeably; both refer  
to a data-carrying entity in a simulation.  The term ``cell'' refers  
to a cubical region of space containing (possibly zero) particles; a  
cell is represented as a node in the tree data structure that stores  
data used by the treecode, and phrases like ``daughter cell'' refer to  
the combined knowledge about a spatial region in the simulation and  
its topological location within the tree.  
  
Conventional implementations of tree structures represent connections  
between nodes within the tree as pointers stored in parent nodes that  
point to the memory locations of the daughters.  This technique is  
difficult to implement on a parallel machine: a parent node on one  
processor may have daughter nodes located on different processors.  
Our treecode instead uses multi-bit ``keys'' to identify particles and  
cells in the tree; a hash table translates keys into actual memory  
locations where cell data is stored (hence the word ``hashed'' in the  
phrase ``parallel hashed oct-tree algorithm'').  This level of  
indirection enables uniform handling of local and non-local cell data;  
the algorithm that maps hash keys to cell data can request non-local  
data from other processors and only make it available to the local  
process once it has arrived.  
  
\subsubsection{Key Construction}  
  
As was mentioned above, each body and cell in the tree is identified  
by a key; by using the key as an index into a hash table, the code can  
quickly locate or request delivery of data about any entity in a  
simulation, and the hash algorithm transparently handles retrieval of  
data stored on other processors in a parallel machine.  The algorithm  
described below generates keys that, when sorted, arrange the  
particles into Morton order within the volume of solution; such keys  
are often called ``Morton keys''.  Morton ordering results in a list  
of particles that fairly well reflects spatial locality of bodies in  
the tree, i.e. particles with Morton keys that fall near each other in  
the sorted list of keys generally lie near each other in the simulated  
volume.  However, particles sorted into e.g. Peano-Hilbert order can  
exhibit an even higher correlation between proximity in the sorted key  
list and proximity in space, which can lead to more efficient  
distribution of data on a parallel computer, so modern treecodes have  
tended to use Peano-Hilbert keys to refer to particles (see, e.g.,  
Springel 2005).  Our treecode supports both Morton and Peano-Hilbert  
keys, but in practice we have found no noticeable difference in  
performance; we generally use Morton keys, as described below.  
  
Each Morton key is a set of $l_k$ bits derived from the $d$ floating  
point coordinates of a body in $d$-dimensional space.  One bit is  
reserved as a placeholder, for reasons explained below, which leaves  
$n = (l_k - 1)/d$ bits to represent each of the coordinates of a body.  
To generate a key for each particle in a simulation, we start by  
calculating the spatial extent of the simulation and then divide the  
largest spatial dimension into $2^n$ equal intervals; the intervals  
can then be indexed by a $n$-bit integer.  For simplicity, the same  
interval spacing is used for the other (smaller) spatial dimensions as  
well.  Each floating point coordinate of each body in the simulation  
is mapped to the $n$-bit integer index of the spatial interval  
containing that coordinate.  
  
To construct a key from the $d$ integers derived from the spatial  
coordinates of a body, the integers are interleaved bit-by-bit.  The  
result is a key composed of $n$ groups of bits, each of length $d$,  
where the bits in the $i$th group are the $i$th bits of each of the  
integers identifying the body, arranged in dimension order.  
  
\subsubsection{Hashing}  
  
The mapping between keys and pointers to cell data is maintained via a  
hash table.  The actual hashing function is very simple---we select  
the $h$ least-significant bits of the key---but because a key reflects  
the spatial location of the associated cell, the spatial distribution  
of bodies and cells determines which hash table entries are filled.  
Each hash table entry is stored in a ``hcell''; the hcell contains a  
pointer to the actual cell data (if the data is local) and maintains  
knowledge of the state of the cell data---whether it is local or  
nonlocal, or if it has been requested from another processor.  
  
\subsubsection{Tree Construction}  
  
The key length $l_k$ determines how many levels of cells the tree can  
represent, or how close two particles can be, relative to the spatial  
extent of the simulation, and still be stored separately in the tree.  
Intermediate (i.e. non-leaf) cells in the tree can be represented in  
the same key space as individual bodies if the highest bit of every  
key is set to 1 as a placeholder; the position of this placeholder  
(the highest non-zero bit) in a cell's key indicates the depth of that  
cell in the tree.  Given a key for any cell or body, the key of the  
parent cell can be found by right-shifting the key by $d$ bits (i.e.  
the number of spatial dimensions).  The root of the tree has a key of  
``1''---the placeholder bit is the lowest (and only) bit in the key.  
  
Keys constructed for bodies are used first to sort the particles and  
distribute them across the set of processors used in a parallel  
calculation (see \S \ref{Code:Parallel}).  After each processor  
receives the bodies assigned to it, a tree of cells is constructed and  
all local bodies are inserted into the tree.  The key associated with  
a body is not changed when the body is added to the tree; instead, the  
body is associated with a cell, and the cell key represents the  
location and depth of the cell in the tree.  The first body is  
inserted into a cell immediately below the root.  Subsequent cells are  
inserted by starting from the location of the previous cell and  
searching for the appropriate location in the tree; because the bodies  
are already sorted, the correct location is usually very close to the  
previously-inserted cell.  Often the insertion of a new cell will  
require a previously-inserted cell to be split, and both the old and  
new cell will be moved to a lower location in the tree; empty cells  
are inserted at each tree level between the old and new locations.  
  
An intermediate cell is ``finished'' when it is clear that no new  
bodies will be inserted below it---because the bodies have already  
been sorted into Morton order, it is easy to determine when no  
additional bodies will be inserted below a particular cell.  The  
process of ``finishing'' includes calculating the total mass and  
spatial extent of the set of all daughter cells, which permits fast  
evaluation of multipole contributions during calculation of  
gravitational forces among groups of particles.  
  
\subsection{Parallelization}\label{Code:Parallel}  
  
Sorting the list of body keys is equivalent to arranging the bodies in  
Morton order.  Morton ordering does a reasonably good job of  
maintaining locality of data in the sorted list; bodies that are close  
together in space end up close to each other in the list.  This  
ordering of bodies also allows easy domain decomposition for  
parallelization; data can be distributed over a set of processors in a  
parallel machine by cutting the list of bodies into ``equal-work''  
lengths and sending each list section to a different processor.  The  
``work'' required to update a particle is usually defined as the  
number of interactions in which the particle participated during the  
previous timestep, which generally results in good load-balancing.  It  
is also possible to adjust the estimated work associated with any set  
of particles to account for, e.g., complex equation of state  
calculations which take a predictable number of iterations to converge.  
  
Parallel tree construction adds one additional step to the process  
described in \S \ref{Code:Tree}; after building a local tree, each  
processor finds and transmits a set of ``branch cells'' to all other  
nodes.  The set of branch cells on a given processor is the coarsest  
set of cells that contains all of the data stored on that processor.  
Branch cells are the highest ``finished'' cells in the local tree; all  
cells and bodies below a branch cell are also stored on the processor,  
while cells above the branch cell include regions of space containing  
particles stored on other processors.  Every processor broadcasts its  
branch cells to every other processor, so that each processor can  
directly request non-local cell data from the correct processor during  
traversals of the tree.  
  
Characterizing the performance of parallel scientific codes is  
difficult, and trying to doing so for SNSPH presents all the usual  
pitfalls.  Per-processor performance metrics, and scaling of  
performance with number of processors in a parallel calculation, both  
vary strongly depending on the size of simulation under consideration,  
the hardware platform on which the simulation is run, and even the  
particular physical or numerical conditions present in the simulation.  
Scaling for SNSPH is linear on hundreds through thousands of  
processors on modern supercomputers, as long as the problem being  
simulated is sufficiently large, and the definition of "large" for a  
given computer depends on the details of the CPU architecture, memory  
subsystem, node interconnects, and other system components.  For a  
core-collapse simulation including all physics modules within SNSPH, a  
4-million-particle set of initial conditions is sufficiently large to  
support nearly linear scaling on up to 256 processors on Pink at Los  
Alamos National Laboratory.  
  
The fraction of time spent on different tasks during a typical  
core-collapse simulation at least identifies the most time-consuming  
portions of SNSPH.  A typical core-collapse simulation spends most of  
its time ($\sim 60$\%) calculation forces and updating flow quantities  
in the innermost loop over individual SPH particles; of this time,  
typically only 10\% is spent in the equation of state or performing  
neutrino transport, but this number can grow dramatically in  
pathological cases when the iterative procedures in the EOS are unable  
to quickly converge to a solution.  $\sim 15$\% of the CPU time is  
spent updating gravitational forces among groups of particles, though  
this percentage can change depending on the desired accuracy for the  
gravitational force calculations.  The treecode accelerates both of  
these calculations, but the gravitational force calculations benefit  
more, since the use of a tree accelerations both identification of  
nearby particles and evaluation of the gravitational forces  
themselves.  Updating quantities for SPH particles benefits only from  
the fast neighbor-finding provided by the treecode.  Calculating keys  
for each particle, performing a parallel sort of the keys across all  
processors in a calculation, and building the tree typically consumes  
$\lesssim 10$\% of the total time in a calculation, and most of this  
time is spent shifting particles between processors.  Overall, delays  
associated with message-passing typically account for $\sim 15--20$\%  
of the time per timestep in a typical calculation on a gigabit  
Ethernet network such as the one used in the Space Simulator at Los  
Alamos National Laboratory.  More sophisticated interconnects, such as  
the Myrinet interconnect used on Pink, reduce this percentage  
dramatically (generally to less than 5\%).  
  
\subsection{Time integration}  
  
  
After the rates of change of all flow variables are calculated via the  
SPH equations, we apply an integration scheme to advance all  
quantities to the next timestep.  To update the specific internal  
energy $u$ of each particle, we use the 2nd-order Adams-Bashforth  
method, a 2nd-order method for 1st-order ODEs:  
\begin{equation}\label{eqn:energy}  
u_{i+1} = u_{i} + \dot{u}_{i} \left({dt}_{i} + \frac{{dt}_{i}^{2}}{2 {dt}_{i-1}}\right) - \dot{u}_{i-1}\left(\frac{{dt}_{i}^{2}}{2 {dt}_{i-1}}\right).  
\end{equation}  
The smoothing length $h$ of each SPH particle is updated using the  
2nd-order Leapfrog method:  
\begin{equation}  
h_{i+1} = h_{i} + {\dot{h}}_{i} \frac{{dt}_{i} + {dt}_{i-1}}{2}.  
\end{equation}  
We update the position $x$ and velocity $\dot{x}$ of each SPH and  
N-body particle using the Press method, a 2nd-order method for  
2nd-order ODEs:  
\begin{equation}\label{eqn:velocity}  
{\dot{x}}_{i+1} = \frac{x_i - x_{i-1}}{{dt}_{i-1}} + {\ddot{x}}_{i} \left({dt}_i + \frac{{dt}_{i-1}}{2}\right),  
\end{equation}  
\begin{equation}\label{eqn:position}  
{x}_{i+1} = {x}_{i} + \left({x}_{i} - {x}_{i-1}\right)\frac{{dt}_{i}}{{dt}_{i-1}} + {\ddot{x}}_{i} \frac{{dt}_{i}\left({dt}_i + {dt}_{i-1}\right)}{2}.  
\end{equation}  
Note that $u_{i+1}$ depends on $u_{i}$ and $u_{i-1}$, and $x_{i+1}$  
depends on $x_i$ and $x_{i-1}$ (not $x_i$ and $\dot{x}_{i}$).  
Equations~\ref{eqn:energy}, \ref{eqn:velocity}, and \ref{eqn:position}  
are not self-starting; when a simulation is started (or restarted from  
an intermediate point), we assume that $u_{i-1} = u_{i}$ and $x_{i-1}  
= x_i - \dot{x}_{i} dt$.  At all other times, $\dot{x}_{i}$ is updated  
only for the benefit of the user; the Press method updates the  
position of each particle using the current and previous position.  
  
The term $(x_{i} - x_{i-1})/{dt}_{i-1}$ in equation~\ref{eqn:velocity}  
can suffer from large floating-point roundoff errors if ${dt}_{i-1}$  
is small, the current and previous positions are nearly equal, and the  
precision of the variables used to store the positions is low; for  
this reason, the code stores current and previous positions in  
double-precision variables.  
  
\section{Code Tests}  
  
The core-collapse supernova problem is so complex that testing one  
piece of physics is not sufficient to ensure that the code will work  
on the entire supernova problem.  As an example, the shock tube,  
Sedov, and Sod problems all essentially test how well a hydrodynamics  
code works on shocks (along with some additional tests of boundary  
conditions), but shock formation is not the only physical process  
relevant to core-collapse.  For example, our rotating models require  
strict conservation of angular momentum and minimal artificial angular  
momentum transport.  We must make sure our tests actually verify our  
code in conditions similar to what we expect in our simulations.  
  
Another issue we must worry about in code-testing is fine-tuning the  
code so that it performs well on a particular simplified test.  This  
sort of fine-tuning occurs fairly often by those modeling the shock  
tube problem.  The shock moves along the grid so one can easily add  
code to model the shock well under these conditions.  But will these  
modifications actually help the code perform when the shock is more  
complex and not parallel to the grid?  The shock tube/Sedov/Sod  
problems should be considered 1st order tests of a code's ability to  
model shocks.  More complex tests are necessary to fully trust a  
numerical algorithm.  
  
There are a number of ways to verify and/or validate a numerical code.  
These two words have specific meaning in the computational community:  
verification refers to testing that a code is solving correctly the  
physical equations it is supposed to be solving.  Validation is  
defined as making sure the physical equations used are the right set  
of equations required for that problem.  Validation is often  
interpreted as any test comparing to experiment.  This is only  
strictly true if the comparison experiment is (or nearly is identical  
to) the problem for which one is validating the code.  Most tests in  
astrophysics are verification tests, and we will focus on these here.  
In astrophysics verification tests include comparisons to analytics  
(good for testing specific pieces of physics), convergence studies,  
and comparisons to other codes.  Code comparison is the only true way  
to test a code's validity on a complex problem.  This mode of code  
verification fails if the codes in the comparison problem get the same  
erroneous result because of different weaknesses in the different  
codes.  Although such coincidencies are not unheard of, they are rare  
in code comparisons we have modeled.  
  
The SNSPH code has already been tested in many of the previous papers  
using this code.  Fryer \& Warren (2002) presented a code-comparison  
test comparing the 3-dimensional results of the newly developed SNSPH  
code to the 2-dimensional results from the code described in Herant et  
al.(1994).  The techniques in both these codes are similar, but the  
codes themselves are very different.  Glaring differences due to  
coding errors would likely have been caught.  We are also conducting a  
detailed comparison of Rayleigh-Taylor and Richtmeyer-Meshkov  
instabilities between SNSPH and the RAGE adaptive mesh refinement  
(Fryer et al. 2006).  We have also run precessing disk calculations  
(Rockefeller et al. 2005c), comparing the results to both analytical  
estimates and past SPH calculations (Nelson \& Papaloizou 2000)  
testing angular momentum transport in the code.  Another test of the  
angular momentum run by Fryer \& Warren (2004) is discussed in more  
detail in \S \label{sec:angm}.  
  
Here we present five additional tests of our code to try to cover some  
of the more important pieces of physics for core-collapse supernovae.  
We use the same code for all five tests and the code was not  
fine-tuned to produce better results for any specific test.  Two of  
these calculations test shocks: a Sedov blast wave and a more complex  
set of shocks using the Galactic Center as an experimental test.  The  
Sedov blast wave has an analytic solution and our code can be tested  
to high precision.  But nature is more complex, and the Galactic  
Center experiment allows us to test how our code models complex shock  
structures, albeit with much less precision than our simple test.  If  
the effects of standing accretion shocks are as critical as some in  
supernova believe (Blondin et al. 2003; Burrows et al. 2005), these  
may be the most important tests for the supernova problem.  Issues in   
the gravity algorithm have also led to many erroneous results in  
core-collapse calculations.  To test gravity in the code, we have run  
an adiabatic collapse calculation and a binary orbit calculation.  The  
binary orbit calculation can also be used to test angular momentum  
conservation.  In the binary calculation subsection, we also describe  
techniques in SPH to test the amount of numerical angular momentum  
transport.  Lastly, we have run a check of our neutrino diffusion  
algorithm.  
  
These tests do not cover all the physics necessary for this  
calculation and it is important to continue to search for ideal tests  
for core-collapse supernovae.  But these five tests provide a basis by  
which we can test a range of the important physics necessary to model  
stellar collapse as well as many other astrophysically-relevant   
problems.  
  
\subsection{Sedov-Taylor Blast Wave}\label{sect:sedov}  
  
The Sedov-Taylor blast wave problem was first discussed by Sedov  
(1982) and Taylor (1950); Sedov ultimately developed an analytic  
solution to the problem.  An amount of energy $E = 1$ is deposited at  
time $t = 0$ into a small volume at the center of a uniform-density,  
low-pressure medium (density $\rho_{0} = 1$ and specific internal  
energy $u = 10^{-5}$ in our test) with a gamma-law equation of state  
\begin{equation}  
P = (\gamma - 1) \rho u.  
\end{equation}  
where $\gamma$, the ratio of specific heats in the medium, is chosen  
to be $1.4$.  The deposited energy heats the gas and drives a  
spherical shock wave outward through the medium.  The radius of the  
blast wave $R$ evolves according to the equation  
\begin{equation}  
R = S(\gamma){t}^{2/5}E^{1/5}{{\rho}_{0}}^{-1/5}  
\end{equation}  
where $t$ is the time elapsed since the explosion and $S(\gamma)$ is a  
function of the ratio of specific heats.  The density, pressure, and  
radial velocity behind the shock evolve self-similarly according to  
\begin{eqnarray}  
\rho/\rho_{0} & = & \psi, \\  
P/P_{0} & = & R^{-3}f_{1}, \\  
v_r & = & R^{-3/2} \phi_{1},  
\end{eqnarray}  
where $\rho_{0}$ and $P_{0}$ are the density and pressure of the  
ambient medium and $\psi$, $f_{1}$, and $\phi_{1}$ are all functions  
of $\eta = r/R$.  The simplicity of the initial conditions and  
the existence of an analytic solution make the Sedov-Taylor problem  
one of the most frequently-used tests of a code's ability to model  
strong shocks.  
  
\begin{figure}  
\plottwo{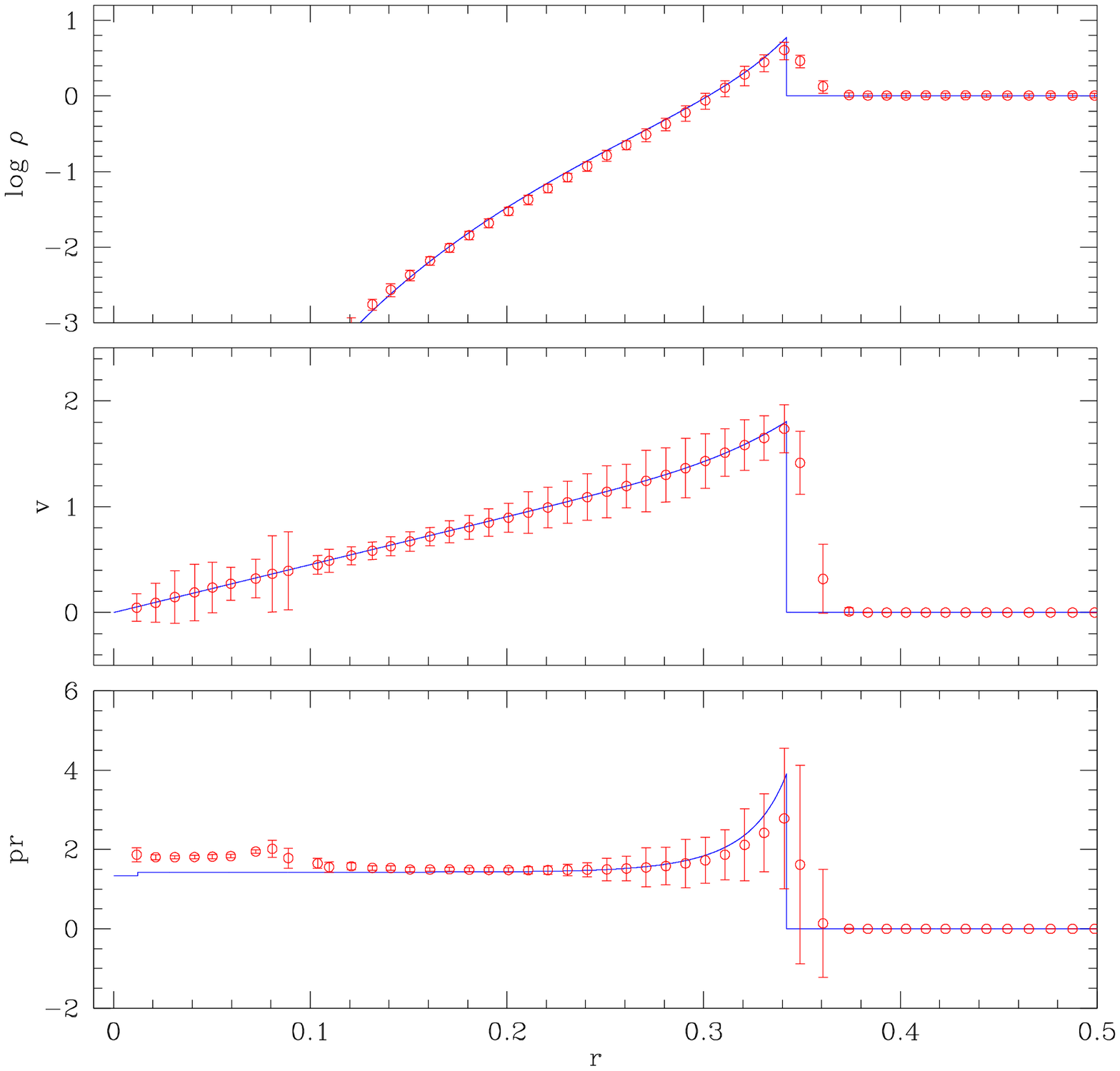}{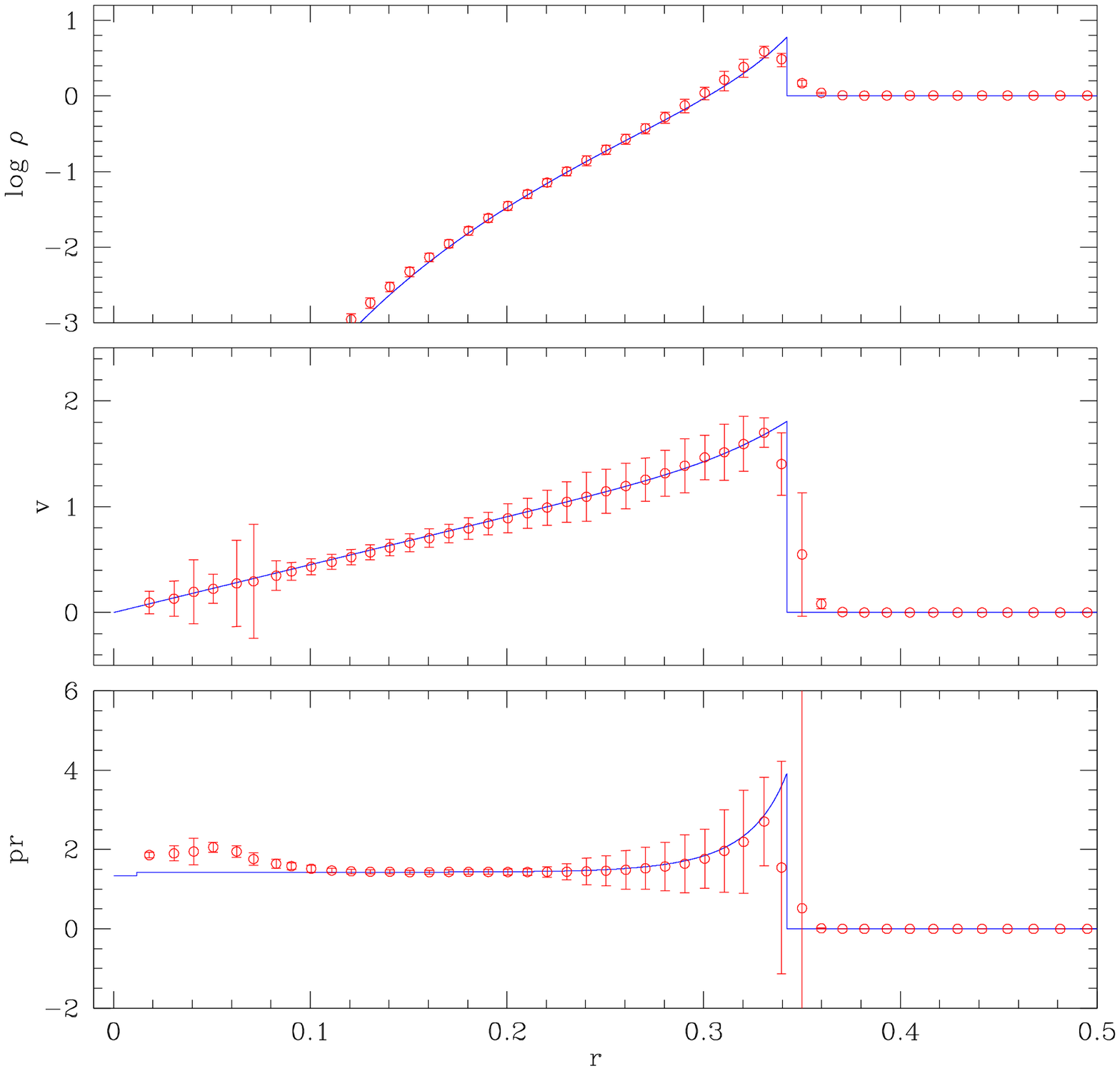}  
  
\caption{Angle-averaged density, velocity, and pressure profiles at time   
  $t = 0.063$ from two SPH simulations of a Sedov blast wave; the blue  
  lines indicate the analytic solution.  The two simulations use  
  slightly different initial conditions, and consequently the results  
  differ slightly as well; for example, the calculated density behind  
  the shock generally falls below the analytic value in the first  
  simulation but above the analytic value in the second.  See \S  
  \ref{sect:sedov} in the text for further discussion.}  
\label{fig:sedov}  
\end{figure}

Figure~\ref{fig:sedov} shows the angle-averaged density, velocity, and  
pressure profiles taken at time $t = 0.063$ from two Sedov explosions  
simulated using SPH.  The two simulations differ only in the specific  
arrangement of particles used for the initial conditions; both sets of  
initial conditions were constructed from concentric spherical shells  
of particles, but the simulation shown in the right-hand plot had an  
inner shell radius ten times smaller than the inner shell radius used  
in the left-hand simulation.  Each circle represents the mass-weighted  
mean flow quantity and location of particles in a single radial bin  
(where the width of each bin is 0.01).  The error bars indicate one  
(mass-weighted) standard deviation around the weighted mean.  The blue  
line indicates the analytic solution.  
  
Part of the scatter in the particle densities and pressures at a given  
radius arises from uncertainties in the initial density profile.  One  
drawback of SPH is that there is an intrinsic scatter in the density  
because of the dependence and variability of the smoothing length.  
Even if we fine-tune the initial model, the scatter will develop after  
a few timesteps.  This scatter places a low-level perturbation to seed  
any convection.  In most past simulations by SNSPH, this scatter had a  
1-sigma error of 10\% in the density.  Although we now can reduce this  
scatter somewhat, it is still an issue with SNSPH calcuations.  In  
stellar collapse, it is likely that the explosive burning just prior  
to collapse produces density perturbations at this level.  If anything,   
the 10\% scatter in our SPH set-up is on par with what we expect to be   
the true initial conditions.  We have run calculations with a factor   
of 2 larger scatter, and it did not alter our convection or mixing   
results in the explosion calculation.  
  
The two simulations also illustrate the effect of different initial  
particle arrangements on flow quantities throughout the simulation.  
Decreasing the radius of the innermost shell of particles results in a  
pressure profile that more accurately matches the analytic solution  
behind the shock front.  Additionally, the density profile in the  
left-hand simulation falls slightly below the analytic value, while  
the density in the right-hand simulation is slightly higher than the  
analytic value.  This variation demonstrates the sensitivity of SPH to  
the initial arrangements of particles in a simulation and emphasizes  
the importance of carefully constructing and testing sets of initial  
conditions.  
  
From these calculations, we see that the SPH calculation produces   
fairly accurate shock velocities, but densities and pressures that   
are low (although the shock entropy is more accurate).  These   
errors decrease with resolution but, in most of our core-collapse   
simulations, we have not yet reached a satisfactory convergence on   
the shock modeling.  A more realistic, but less standard, test   
would have a shock traverse a density gradient instead of a   
constant density profile.  Fryer et al. (2006) compares the   
results of such a test for a number of coding techniques, including   
this SNSPH code.  
  
\subsection{Galactic Center Shocks}  
  
The Sedov blast wave tests a single, spherically-symmetric strong  
shock.  In a grid code, such a test calculation can be easily  
fine-tuned by using a spherically symmetric grid and using solvers  
that assume that the shock front will be parallel to the grid.  
Unfortunately, the supernova problem does not have such a well-defined  
shock and fine-tuning the solver for a parallel grid may well give a  
worse answer for the specific case of stellar collapse.  
  
Devising a test for more chaotic shocks is difficult.  Few have  
analytic solutions.  Instead we present here an experimental test.  
The X-ray emission in the Galactic center has been recently studied in  
detail by the Chandra X-ray Observatory.  This X-ray emission is  
dominated by a point source Sgr A$^{*}$, believed to be a $3.7 \times  
10^6$\,M$_\odot$ black hole (see Ghez et al. 2005, although Rockefeller  
et al. 2004 used a $2.6 \times 10^6$\,M$_\odot$ black hole) accreting  
the gas around it.  But there is also a diffuse X-ray component which  
is believed to be produced by colliding gas in the Galactic center.  
This gas arises from stellar winds from 25 wind-producing stars.  
Since the X-ray emission is proportional to the square of the density  
of the gas, it is an ideal probe of the shocks in this system.  As  
with any experimental test, there are uncertainies both in the initial  
conditions, the relevant physics, and the final measurement.  Let's  
consider both of these sets of uncertainties.  
  
The initial conditions for this problem consist of 25 mass-losing  
stars (the contribution by the other stars near the Galactic center is  
negligible).  Indeed, most of the wind matter arises from the 7  
strongest wind sources.  Table \ref{tab:gc} (from Rockefeller et  
al. 2004) gives the positions, wind velocities, and mass-loss rates  
for all of these stars.  Although the x and y (projected) positions  
are fairly well-known, the z (radial) positions are completely  
unknown.  Rockefeller et al.  (2004) found that taking extreme  
positions for the coordinates of these wind sources led to 15\%  
variation in the X-ray flux.  The other uncertainties lie in the wind  
velocities and mass-loss rates of each of these stars.  Those stars  
for which we have observations of the He I line emission have wind  
velocities that are known, but for many stars, the wind velocities   
and their associated mass-loss rates are merely estimated (see  
Rockefeller et al. 2004 for details).  However, the total mass lost in  
winds as well as the wind structure of the 7 dominant mass-losing  
stars are better constrained, so despite uncertainties in some of the  
stars, these initial wind conditions appear sufficiently constrained  
to use this experiment.  And with time, these initial conditions will  
only become better known\footnote{Rapid variation in the mass loss  
will not affect the X-ray emission.  Measuring the long-term average  
mass-loss from each star is all that is necessary for this experiment,  
allowing astronomers hundreds of years to better constrain the initial  
conditions.}.  
  
The relevant physics for this problem is fairly straightforward.  The  
equation of state is accurately modeled by an ideal gas with a  
$\gamma=5/3$ power law.  The gravitational force is dominated by the  
central black hole (see Rockefeller et al.  2004 for details).  It is  
unlikely that global magnetic fields will be sufficiently strong to  
play a role in the diffuse X-ray emission (although they may effect  
the X-ray emission from the point source).  At the temperatures and  
densities that these shocks produce, the dominant components of the  
continuum emissivity are electron-ion ($\epsilon_{ei}$) and  
electron-electron ($\epsilon_{ee}$) bremsstrahlung (Rockefeller et  
al. 2004 used the standard simplified representations for these  
emission processes).  The only other physics that need be considered  
is the possibility of surrounding molecular clouds which confine the  
wind material.  But these clouds mostly affect the spatial  
distribution of the X-ray emission.  This effect is limited to the  
low-level contours of the X-ray emission, and hence does not effect  
the total X-ray emission considerably.  
  
The strongest observational constraint we can currently apply from the  
Galactic center is the total diffuse X-ray luminosity.  Observations  
currently place this luminosity at $(7.6^{+2.6}_{-1.9})\times10^{31}  
{\rm ergs \, s^{-1} \, arcsec^{-2}}$.  In principal, the X-ray  
contours can be used as a constraint, but the low-level contours are  
very sensitive to the positions of the stars and the structure of the  
surrounding gas.  At this time, more detailed tests of this  
multi-shock problem are best limited to code comparisons.  We provide  
3-dimensional density and energy plots on our website  
(http://qso.lanl.gov/$\sim$clf/codetest) for comparison.  
  
\subsection{Adiabatic Collapse}  
  
The adiabatic collapse of an initially isothermal spherical cloud of  
gas has been used in several investigations of SPH codes with gravity  
(Steinmetz \& M\"uller 1993; Thacker {et~al.} 2000; Springel {et~al.}  
2001, Wadsley {et~al.}, 2004).  We follow the units used in the first  
presentation of this problem (Evrard, 1988).  With $ G = M = R = 1 $,  
the initial density distribution is:  
\begin{equation}  
\rho(r) = M/(2 \pi R^2) r^{-1} = (2 \pi r)^{-1}.  
\end{equation}  
where $M$ is the total mass of the system within the cut-off radius  
$R$.  
  
The initial internal energy of the system is chosen as $0.05 G M/R$,  
with the adiabatic index $ \gamma = 5/3 $.  The initial physical  
density distribution was applied to a grid with hexagonal symmetry  
using the technique of Davies, Benz \& Hills (1992).  The  
gravitational potential was smoothed with a Plummer softening of 0.01  
R (Plummer 1911).  The SPH kernel was initially set to give roughly 60  
neighbors for each particle, and the SPH smoothing length was allowed  
to evolve within the constraints of $h_{max} = 150$ and $h_{min} =  
30$.  
  
\begin{figure}  
\plotone{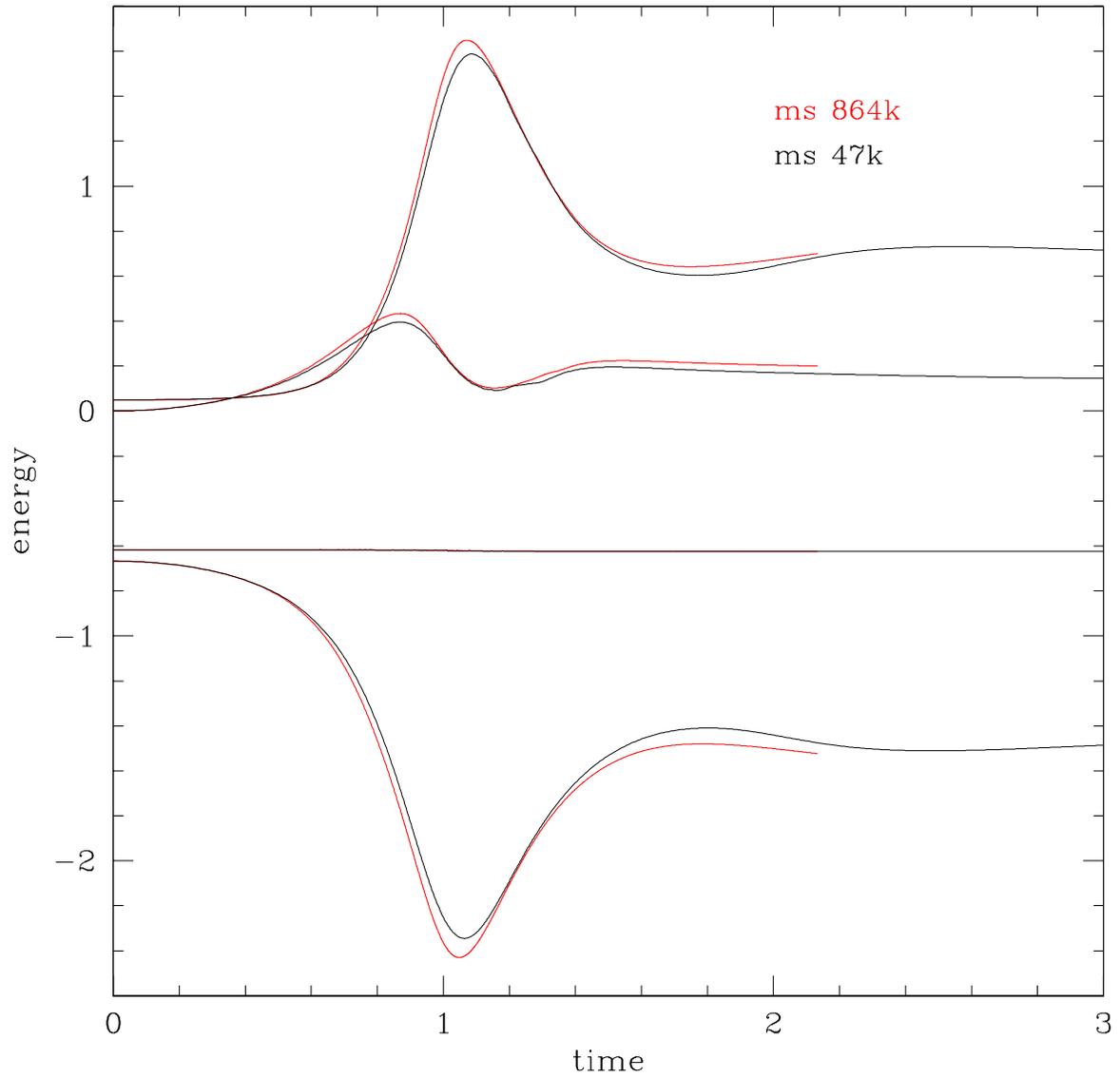}  
\caption{The time evolution of thermal, kinetic, total and  
gravitational potential energies during the adiabatic collapse of an  
initially isothermal gas sphere for SPH calculations with 43,000 and  
864,000 particles.  This figure may be compared with the 1-dimensional  
PPM and 3-dimensional SPH calculations from Figure 6 of Steinmetz \&  
M\"uller (1993).}  
\label{fig:t_vs_e}  
\end{figure}

We show results for simulations with 47,000 particles, and with  
864,000 particles in Figure~\ref{fig:t_vs_e}.  The 864k simulation  
conserves total (kinetic + internal + potential) energy to the 1\%  
level at t=2.1 using 1475 timesteps.  The 47k simulation conserves  
total energy to 1.2\% at t=3.0 using 600 timesteps.  Our version of  
SPH gives very similar results to the version of SPH presented in  
Steinmetz \& M\"uller (1993).  As we increase resolution, our solution  
approaches the solution from their 1-dimensional PPM simulation which  
has a much higher effective resolution per dimension.  
  
\subsection{Binary Orbits: Testing Angular Momentum Conservation}  
\label{sec:angmtest}  
  
One of the true strengths of the SPH method is that it can conserve  
both linear and angular momentum at the same time.  This feature makes  
SPH an ideal technique in modeling binary interactions and rotation in  
core-collapse calculations.  However, even though SPH conserves  
angular momentum between particle interactions and hence is globally   
conserved, a code may still numerically transport that angular  
momentum and one must test both the conservation and artificial  
transport before running simulations that depend heavily on angular  
momentum.  In this subsection, we will focus on a test to determine  
the level at which angular momentum is conserved in our SPH code.  We  
first show how the SPH equations lead to angular and linear momentum  
conservation.  To test the actual application of these equations, we  
follow the evolution of a close binary system for over 15 orbits.  
This test allows an ideal measurement of the angular momentum  
conservation in a system.  We end with a discussion of methods to test  
the importance of numerical angular momentum transport.  
  
\begin{figure}  
\epsscale{.8}  
\plotone{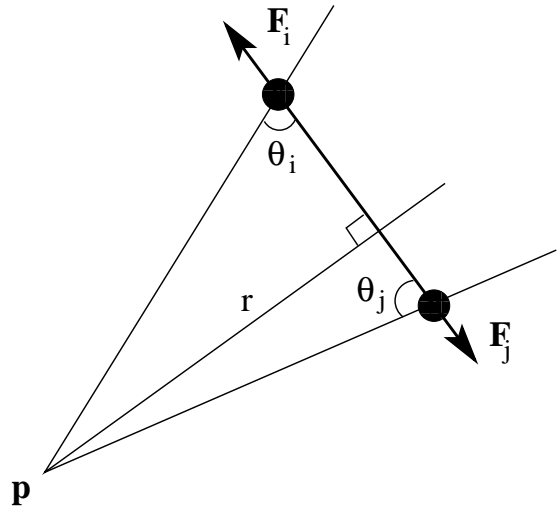}  
  
\caption{Diagram of the force interactions between a pair of particles  
in an SPH calculation.  The torque on particle i from particle j  
viewed from an arbitrary reference point p is $(r/sin\theta_i)F_i  
sin\theta_i$.  Using $F_i=-F_j$ from \S 2, one can easily show that   
the torque on particle $i$ from particle $j$ is equal, but opposite,   
to the torque exerted on particle $j$ by particle $i$.}  
\label{fig:pair}  
\end{figure}

Angular momentum is conserved by noting that the forces are always  
directed along a line joining the particle pairs  
(eq. \ref{eq:momcons}).  Recall that the torque on a given particle  
$i$ by particle j is $\tau_{i,j} = \overrightarrow{r_i} \times  
\overrightarrow{F_i}$ where $\overrightarrow{r_i}$ is the vector from  
a reference point to particle $i$ and $\overrightarrow{F_i}$ is the  
force on particle $i$ due to particle $j$.  Figure \ref{fig:pair}  
shows our particle pair.  Because the acceleration of our particles is  
along the line joining these particles, the torque on any two  
particles are equal in magnitude, but opposite in sign, balancing each  
other and leading to angular momentum conservation:  
\begin{equation}  
\tau_i= r/sin(\theta_i) F_i sin(\theta_i) = r F_i = r/sin(\theta_j) F_j sin(\theta_j) = -r F_i = -\tau_i,  
\end{equation}  
where we have used from equation \ref{eq:momcons} the relation $F_i =  
m_i ( d \overrightarrow{v}_i/dt )_j = - m_j ( d  
\overrightarrow{v}_j/dt )_i = - F_j$.  The SPH equations strictly  
conserve linear and angular momentum.  
  
To test how well such a formulism works in an applied problem, we  
consider the problem of a binary system in an extremely close orbit.  
We use two equal-massed stars with a semi-major axis of $2 \times  
R_{\rm star}$.  The initial stars were evolved for the equivalent of 6  
orbits as single stars before being placed into this binary system.  
These two single stars were then placed in a close orbit around each  
other with spin periods set to the orbital period.  This close orbit  
was chosen to study the angular momentum conservation in an extreme  
situation, but bear in mind that the stellar radius is roughly 20\%  
larger than its Roche-radius, so as time proceeds, the stars will  
gradually lose matter.  But since we run our simulations only for 18  
orbits, we shall see that this is only a large effect at late times.  
Such a binary test simulation, but under less extreme orbital  
conditions, has been done with grid-based codes (e.g. Motl, Tohline,  
\& Frank 2002) and we compare, where possible, with these simulations.  
  
\begin{figure}  
\epsscale{1.0}  
\plotone{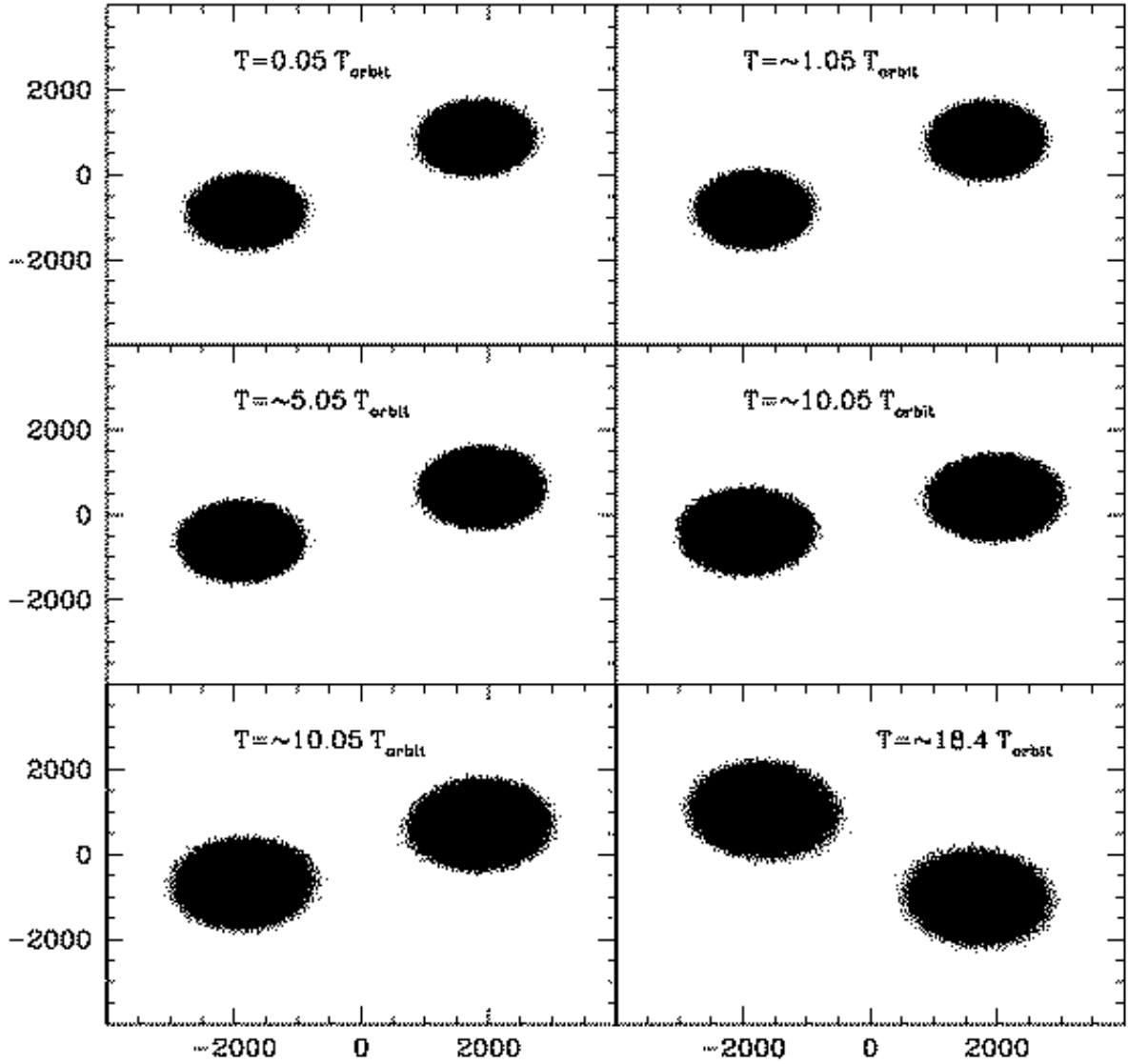}  
  
\caption{Snapshots in time of the evolution of a binary system put  
just inside the Roche-radius separation.  The particles shown are  
limited to a slice in the orbital plane, $-50<z<50$ in code units.  
Although most aspects of this problem are scalable, the actual problem  
is of two 16\,M$_\odot$ stars with a separation of $4\times10^{11}$cm  
placed in a near circular orbit.}  
\label{fig:partp}  
\end{figure}

\begin{figure}  
\plotone{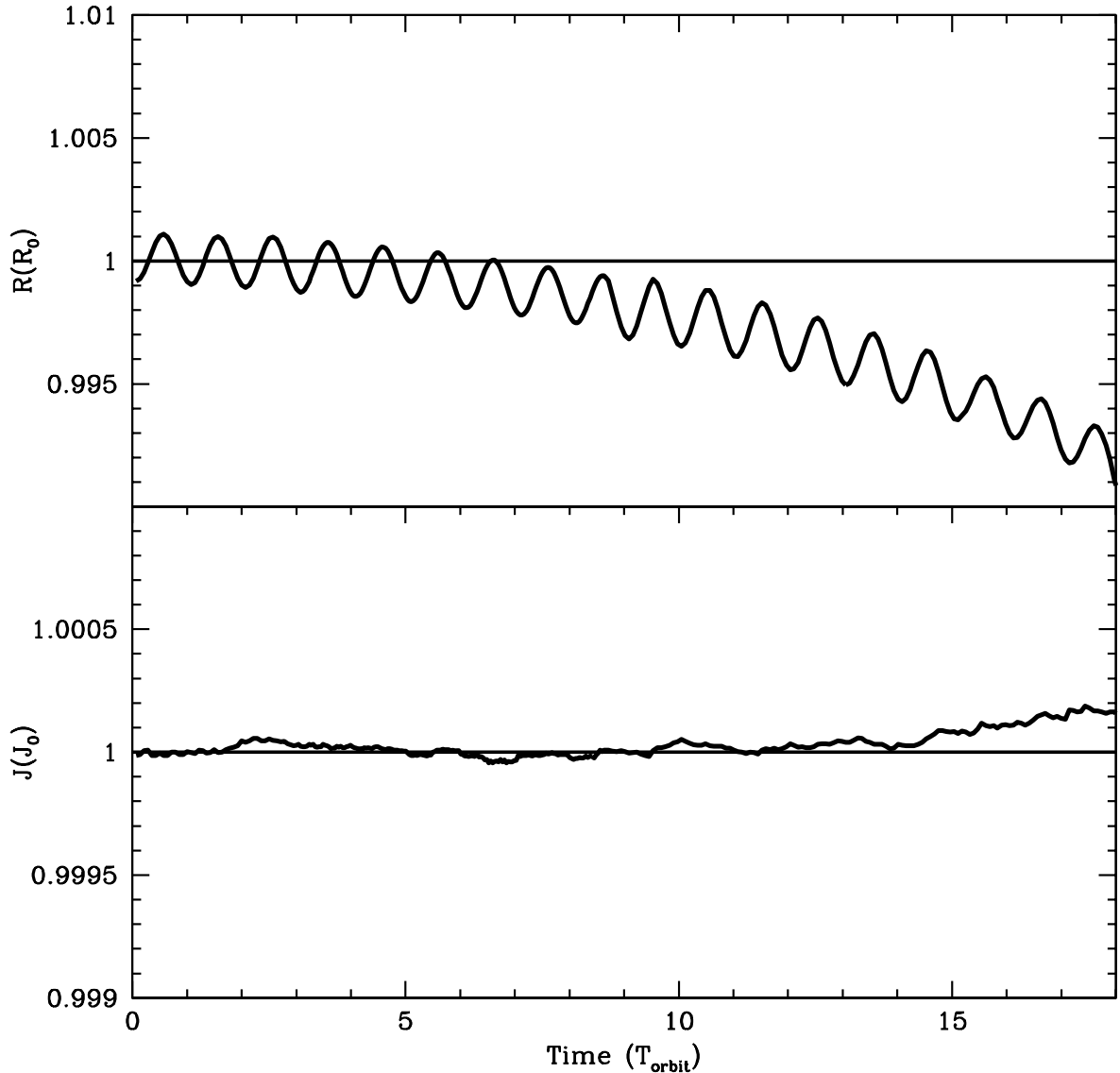}  
  
\caption{Top: Mass-averaged orbital separation (in units of the  
initial separation) versus time (in units of orbital periods) for the  
binary simulation.  The oscillations arise from errors in the initial  
orbital velocities.  As mass begins to fill the Lagrange point, the  
mass-averaged orbital separation begins to decrease.  Bottom:  total   
angular momentum (in units of initial total angular momentum) versus   
orbital time.  The total angular momentum is conserved to better than   
0.01\% for 15 orbits.}  
\label{fig:angm}  
\end{figure}

Fig. \ref{fig:partp} shows a slice ($|z|<50$ code units) in particle  
distribution for our binary system at 6 different times (roughly  
corresponding to the same point in the orbit except for the last time  
which corresponds to our final time dump).  No particles leave the  
system (all particles remain bound to their star).  However, because  
they are both overfilling their Roche-radius, the stars slowly expand  
with time, preferentially building up mass at the L1 Lagrange point.  
Fig. \ref{fig:angm} shows the angular momentum as a function of time  
(bottom) and the orbital radius as a function of time (top) for the  
binary.  The orbital radius oscillates because of a slight error in  
our choice of the orbital velocity in the initial condition.  As we  
shall see, this ultimately causes the code to break.  As we are using  
our tests to find weaknesses in the code, we believe this initial  
deviation from a circular orbit a nice additional test to the code.  
In addition, the orbital serpartion drops by nearly 1\% after 18  
orbits.  This is mostly because material is piling up at the L1 point,  
causing the mass-weighted center of the stars to move slightly toward  
that Lagrange point.  The orbital angular momentum remains much better  
conserved, deviating by only 0.01\% after 15 orbits.  This occurred  
with fairly lenient error tolerances on our MAC and no tuning of the  
code to address this problem.  The lack of angular momentum conservation   
is a result of the distortion in the stars coupled to the lenient tolerances   
on the MAC (leading to growing errors in the gravitational acceleration).    
A circular orbit initial condition, or more stringent MAC tolerances   
could reduce this error.  In comparison, the tuned simulations  
(the best we have seen thusfar by grid codes) by Motl et al. (2002)  
using grid methods found deviations of nearly 0.08\% after only 5  
orbits!  
  
\begin{figure}  
\plotone{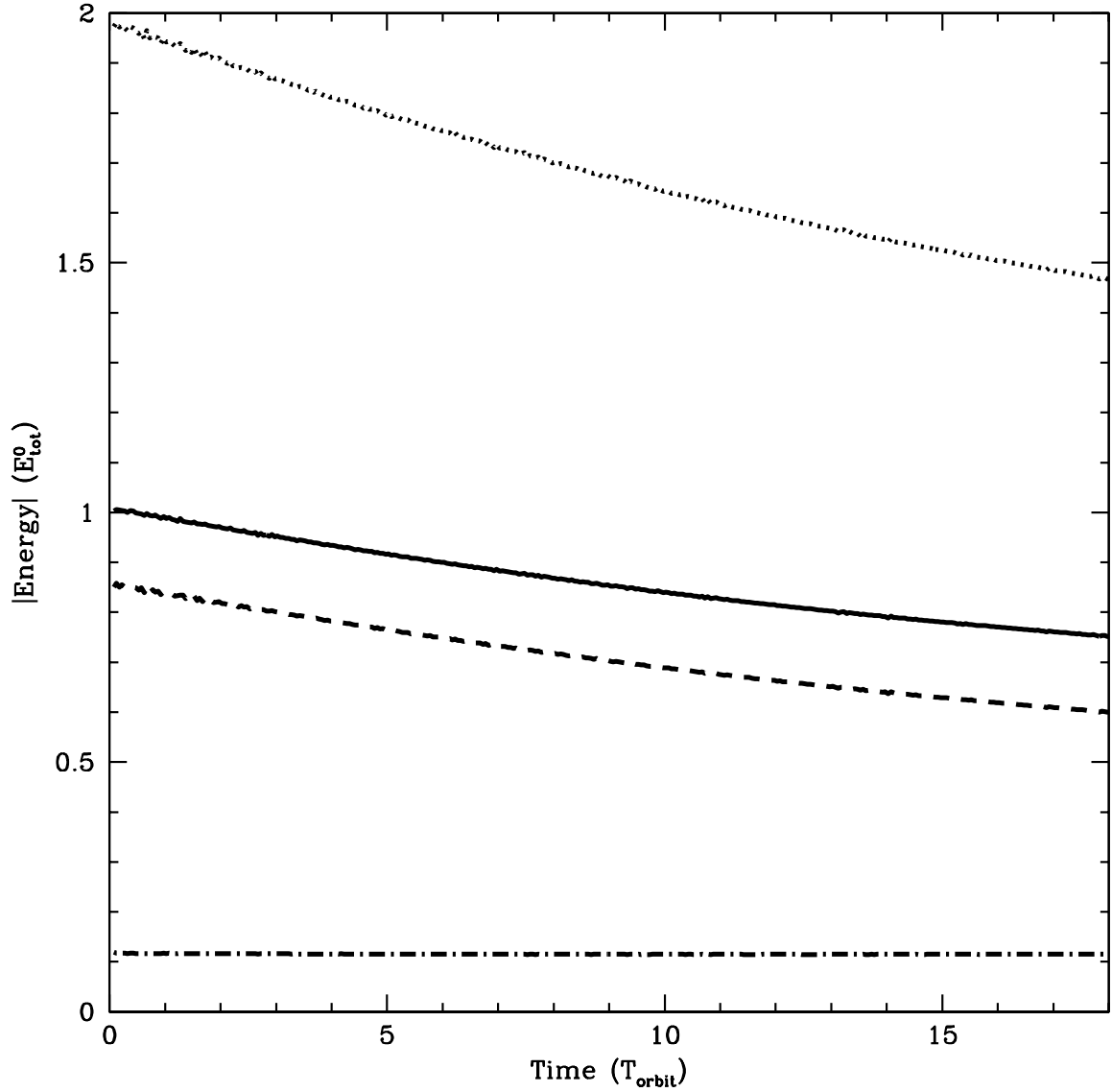}  
  
\caption{Absolute energy versus orbital time.  The total  
(kinetic+thermal+potential) energy (solid line) and potential energy  
(dotted line) are both negative (the system is bound).  Artificial  
viscosity causes the star to heat up, but because it then expands, the  
total thermal energy (dashed line) actually decreases with time.  The  
kinetic energy (dot-dashed line), however, remains relatively  
constant, a reflection of the angular momentum conservation.}  
\label{fig:energy}  
\end{figure}

But not all of the expansion of the star is due to Roche-Lobe  
overflow.  The tidal forces in this problem lead to friction that,  
with our artificial viscosity, causes the star to heat up and expand.  
This numerical artifact of codes with artificial viscosity leads to  
poor energy conservation.  Many fixes exist (Balsara 1995; Owen 2004),  
but in most of the problems we have study, shocks develop quickly and  
are fairly extensive.  For such problems with widespread shocks, these  
fixes do not seem to make a noticeable difference in the simulation  
and we do not include these techniques in any of the current simulations done  
with this code.  Figure \ref{fig:energy} shows the absolute value of  
the energy components as a function of time in units of the total  
energy.  The potential (dotted line) and, because the stars are bound,  
the total (solid) energies are negative.  The magnitude of the total,  
potential, and thermal energy all decrease with time because of the  
expansion of the stars.  The kinetic energy, the primary diagnostic of  
the orbits, remains relatively constant (which is a simple reflection  
of the conserved angular momentum and the rough conservation in the  
orbital radius).  Because of the high artificial viscosity in this  
simulation, the system gains 10\% of its total energy after 10 orbits,  
and another 15\% after 18 orbits.  The artificial viscosity terms were  
reduced slightly from our standard set: $\alpha=1.0,\beta=2.0$, but  
varying this value did not change our momentum conservation  
noticeably.  At the expense of shock modeling, we could decrease the  
artificial viscosity to reduce the errors in the energy conservation.  
  
\begin{figure}  
\plotone{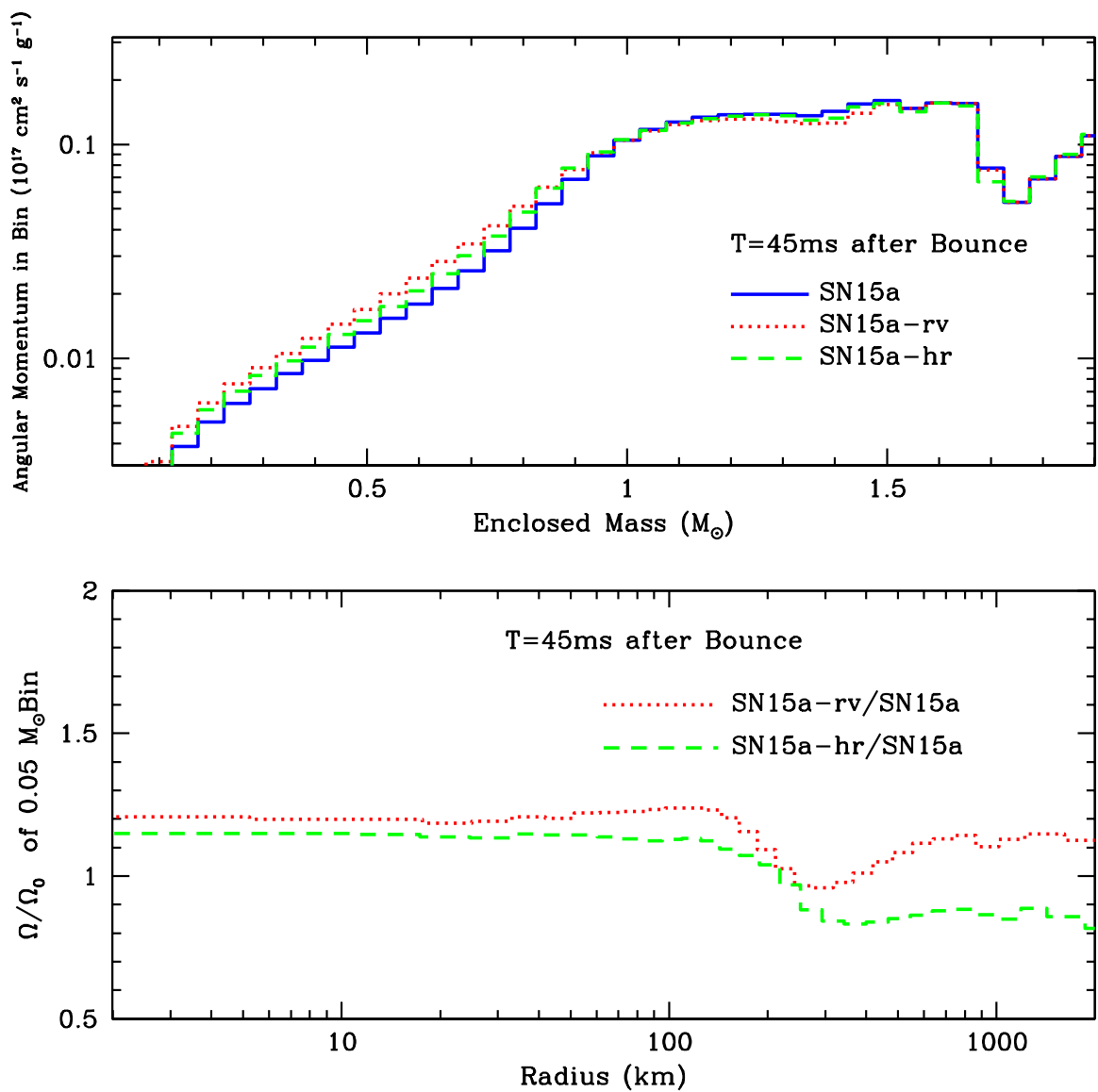}  
  
\caption{Top: Angular momentum versus enclosed mass for a rotating  
collapsing star (Fryer \& Warren 2004).  The 3 lines represent the  
standard model (solid line), a model using an artificial viscosity  
reduced by a factor of 10 (dotted line), and a high resolution model  
(dashed line).  Bottom: The ratio of these angular velocities: reduced  
viscosity/standard (dotted line) and high resolution/standard  
(dashed).  The answer does not change by more than 10-20\% when  
reducing the artificial viscosity, and it is unlikely that we are off  
by more than that value for the angular momentum in our simulations.}  
\label{fig:angtra}  
\end{figure}

The artificial viscosity in SPH, acting like any viscosity (real or  
numerical), leads also to angular momentum transport.  Fryer \& Warren  
(2004) worried about this specific effect in their collapse  
calculations of rotating supernovae.  Fortunately, numerical viscosity  
can be controlled and its effects can be understood.  Figure  
\ref{fig:angtra} shows the angular momentum profile versus mass (top)  
for a standard model (1 million particles - solid line), a model with  
high resolution (5 million particles - dashed line) and a module with  
the artificial viscosity reduced by a factor of 10 (dotted line).  The  
high resolution and reduced viscosity simulations reduce the effect of  
numerical angular momentum transport (in the reduced viscosity  
simulation, by nearly a factor of 10).  The bottom panel of figure  
\ref{fig:angtra} shows the ratio of angular velocities of the two  
modified simulations with respect to the standard simulation.  What we  
find is roughly 20\% more angular momentum in the core.  With our  
extreme reduction in viscosity, this corresponds to the maximum errors  
in the angular momentum transport caused by numerical viscosity.  This  
result eased the concerns Fryer \& Warren (2004) had regarding the  
effects of numerical viscosity on the angular momentum transport in  
their collapse calculations.  
  
Both the conservation and numerical transport of angular momentum must  
be studied when modeling rotating systems.  In most cases, SNSPH  
exhibits very few numerical errors with angular momentum and is an  
ideal technique for rotating problems.  
  
\subsection{Flux-Limited Diffusion Test}  
  
The flux-limited diffusion scheme described in \S \ref{Code:RadTran}  
follows the radiation diffusion between a particle and all of its SPH  
neighbors.  This scheme, originally developed by Herant et al.  
(1994), uses the neighbor list generated to evaluate SPH forces and  
can be parallelized using the same techniques used to parallelize SPH.  
Its major limitation on the scalability arises from converting from  
the flux-limited diffusion algorithm to the free-streaming solution.  
At this point, a global sum must be made over all processors, placing  
a synchronization point in the code.  But, in general, it is both fast  
and easy to implement.    
  
Testing radiation transport schemes in general would, and has,  
comprised many papers in itself.  Here we focus on a simple test  
comparing the SPH flux-limited scheme to a 1-dimensional grid-based  
flux-limited transport scheme.  This test does not prove the  
applicability of flux-limited diffusion to the supernova problem, but  
it does show that our scheme for putting flux-limited diffusion into  
SPH does work.  
  
For the initial conditions of this test, we use a  
spherically-symmetric neutron star atmosphere (material below  
$10^{14}$\,g\,cm$^{-3}$) for a neutron star roughly 130\,ms after  
bounce.  We map this structure onto our SPH particles, setting up   
the particles in a series of spherical shells.  The shells are   
produced by placing particles randomly on a sphere and using   
repulsive forces to evenly separate these particles within   
a predetermined tolerance.  We make a similar setup  
in a 1-dimensional grid using one zone per shell of SPH particles.  With  
this setup, we minimize the differences between the density and  
temperature structure of our 1- and 3-dimensional models.  We  
determine the trapping radius to be 25\,km.  Below this radius, we set  
the electron neutrino fraction ($Y_{\nu_e}$) to 0.15 with a mean  
energy of 10\,MeV.  Above this radius, $Y_{\nu_e}$ is set to zero.  
  
\begin{figure}  
\plotone{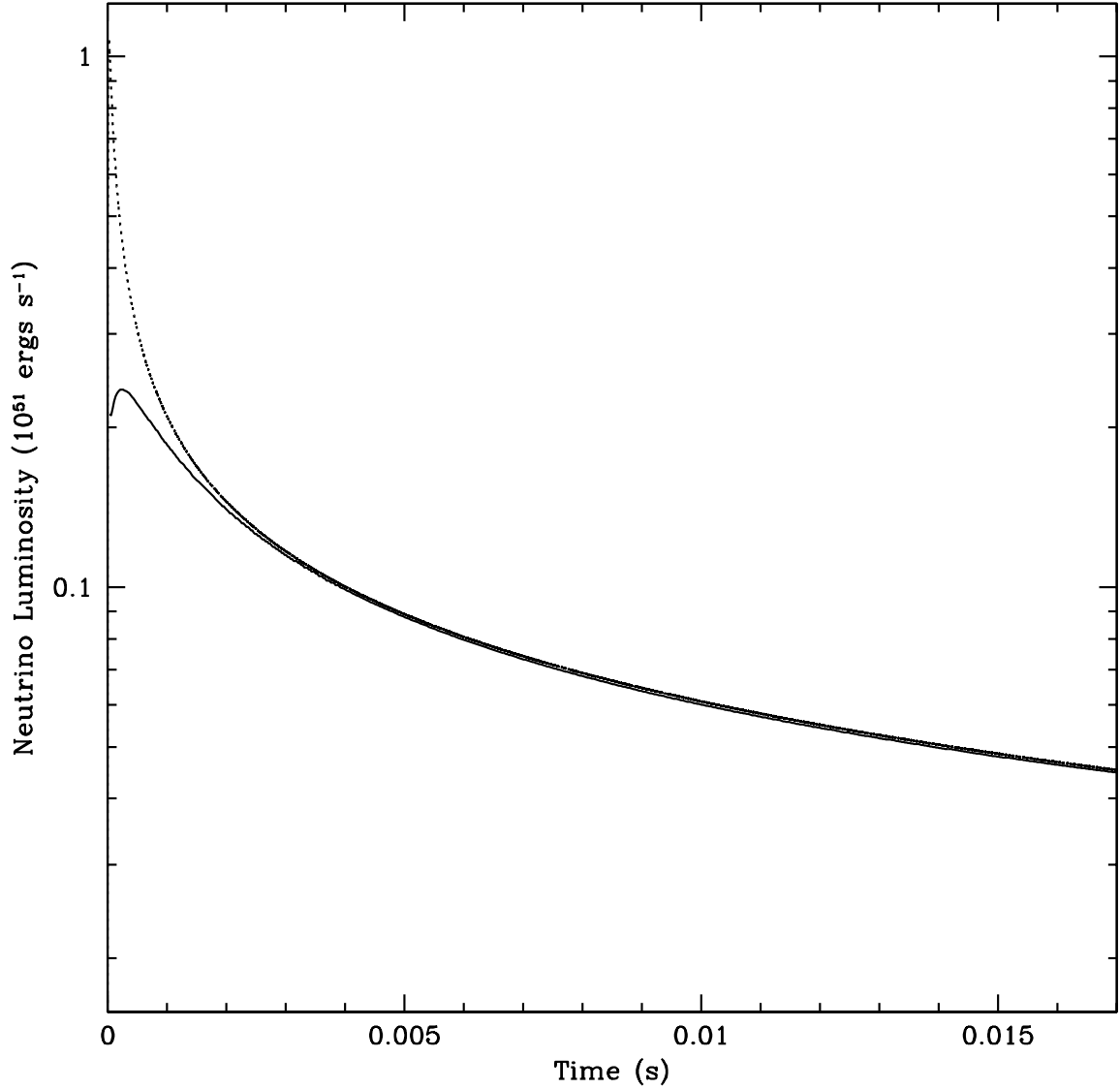}  
  
\caption{Neutrino luminosity as a function of time for our test of the   
  flux-limited diffusion algorithm in SNSPH.  The dotted line shows  
  the results using SNSPH and the solid line is the result from the  
  1-dimensional flux-limited diffusion code.  Due to the smoothed  
  nature of the algorithm in SPH, there is an initial spike in the SPH  
  result.  However, after the first few milliseconds, the two results  
  converge and agree to less than a percent.}  
\label{fig:nulum}  
\end{figure}

Our test focuses on the neutrino transport alone; we evolve only the  
electron neutrino fraction with time and hold the density,  
temperature, and electron fraction fixed.  We allow no new neutrino  
emission.  The flux arising from our neutrino trapping radius versus  
time for both our 1-dimensional grid simulation and our 3-dimensional  
SPH calculation is plotted in figure \ref{fig:nulum}.  The initial  
flux of the SPH calculation is higher, since the neighbors of any  
particle extend beyond the equivalent of an adjacent cell for the  
1-dimensional calculation.  In general, the luminosity for both  
these calculations agree to better than 3\%.  
  
More detailed tests are required to truly trust this transport   
scheme.  Sufficiently complex tests can not be solved analytically   
and we are currently developing a set of comparison calculations   
using Monte Carlo (with its well-defined errors) as a solution   
(Hungerford et al. 2006).  
  
\section{Conclusions}  
  
The SNSPH code is built upon the parallel tree code of Warren \&  
Salmon (1993,1995).  This code has proven extremely portable and  
scalable (beyond 1000 processors) on a number of computer  
architectures from large supercomputers --- the Advanced Simulation and  
Computing Program (e.g. Red, Pink, Q) and National Energy Research  
Scientific Computing Center (e.g. Seaborg) --- to local Beowulf clusters  
--- e.g. Space Simulator ---.  SNSPH uses the Benz version (Benz 1989)  
of Smooth Particle Hydrodynamics to model the Euler equations. A  
number of equations of state have been implemented to close these  
equations, from ideal and perfect gas equations of state to planetary  
equations of state (Benz, Slattery, \& Cameron 1986) to the equation  
of state cocktail used to model core-collapse (Herant et al. 1994).  
SNSPH also includes an explicit flux-limited diffusion scheme to  
moderate neutrino transport (Herant et al. 1994, Fryer \& Warren  
2002).  This transport scheme is even now being modified to model  
photon transport.  
  
Although the focus of SNSPH has been to study stellar collapse (Fryer  
\& Warren 2002,2004; Fryer 2004) it has also been adapted to model  
stellar explosions (Hungerford et al. 2003, 2005) and the Galactic center  
(Rockefeller et al. 2004, 2005) with a list of ongoing projects that  
spreads even further to planet formation and cosmology.  
  
But, as with any computational technique, this code has both  
advantages and disadvantages.  We have presented a number of tests   
of this SNSPH code.  These test problems go beyond the standard  
``shock'' tests (Sedov, Sod problem, shock tube) common in most test  
suites and include tests of the broad range of physics modelled in  
core-collapse.   Our tests mark just the beginning set of tests  
SNSPH must pass if it is to become a multi-purpose radiation  
hydrodynamics code.  
  
Even so, we can already outline the strengths and weaknesses of the  
SNSPH technique.  The particle based scheme allows modellers to  
preserve many advantages of a Lagrangian scheme in turbulent flows  
where typical grid-based Lagrangian cells become tangled.  We have  
shown that the hydrodynamics scheme conserves both total energy and  
momentum.  Our scheme is easily modified to include a wide variety of  
equations of state and external forces and the transport scheme can be  
modified to model the transport of photons as well as neutrinos.  The  
advantages include:  
\begin{itemize}  
\item[1:] Lagrangian technique allows the resolution to follow the  
mass.  This is ideal for problems that cover a large range of  
spatial scales, but that focus the mass in one place (e.g.  
supernova explosions).  It is also ideal for problems where  
the area of interest moves with time  (e.g. in neutron star  
kicks and in binaries).  
\item[2:] The Lagrangian scheme also avoids any numerical diffusion  
(either heat or elemental abundances) that plague any Euler  
scheme during the advection step.  Such artificial diffusion  
can lead to disastrous results in problems where the abundances  
must be tracked exactly (e.g. nucleosynthesis issues in supernova  
explosions) and abundance gradients are large.  The same is  
true for problems where the temperature gradient is large.  
\item[3:] SPH conserves angular momentum and linear momentum at  
the same time.  
\item[4:] The explicit flux-limited diffusion transport scheme scales well  
(for transport schemes).  
\end{itemize}  
The disadvantages of the particle based method are few and can be  
summed up in three limitations with respect to core-collapse supernovae:  
\begin{itemize}  
\item[1:] SPH initial conditions tend to have small perturbations on the  
5-10\% level that will artificially seed convection.  Such  
perturbations are difficult to reduce completely and their effects  
must be understood when interpreting simulation results.  
\item[2:] The artificial viscosity term added to model shocks does not  
work as well as Riemann methods when the shock moves along the grid.  
However, when the shock is not so well behaved, we believe (and  
propose a test for comparison) that SPH behaves as well as most grid  
codes.  The artificial viscosity also leads to both an artificial  
angular momentum and artificial heating term.  In shear flows, such  
numerical effects can lead to unrealistic heating.  We have shown  
(Fryer \& Warren 2004) that this is not an issue for even our  
fast-rotating supernova cores.  
\item[3:] Transport schemes are generally designed for grid-based codes.  
The flux-limited transport scheme in this code works at a basic level,  
but more sophisticated schemes have not been incorporated into SPH,  
and much more work must be done to prove that more sophisticated  
schemes can be added smoothly.  
\end{itemize}  
  
No code is ideal for all problems.  Without the development of a more  
sophisticated transport scheme for SPH (whether it be a direct  
discretization method such as $S_n$ or a monte-carlo technique), our  
SNSPH code will not be able to model the detailed neutrino transport  
many believe is necessary to get a final solution to the core-collapse  
supernova problem.  New transport schemes applicable for SPH are being  
pursued and the current dearth of schemes may not prove a long-term  
limitation.  There are many problems that are most easily solved with  
a Lagrangian technique: from supernova explosions to binary mergers to  
galaxy collisions.  SNSPH is ideally suited for these problems and has  
the versatility to adapt to these problems quickly.  With other  
problems, such as black hole accretion disks, the gas accretion during  
the formation of Jovian planets or the complex shock structure in the  
Galactic center, SNSPH has its advantages and disadvantages over  
grid-based codes.  SNSPH is a powerful tool in developing a solution  
to these outstanding astrophysical problems.  
  
{\bf Acknowledgments} This work was funded under the auspices of the  
U.S.\ Dept.\ of Energy, and supported by its contract W-7405-ENG-36 to  
Los Alamos National Laboratory, by a DOE SciDAC grant number  
DE-FC02-01ER41176 and by NASA Grant SWIF03-0047-0037.  The simulations  
were conducted on the Space Simulator and ASC PINK at Los Alamos National  
Laboratory.  
  
{}  
  
\clearpage  
  
\begin{deluxetable}{lcccccc}  
\tablewidth{0pt}  
\tablecaption{Parameters for Galactic Center Wind Sources\label{tab:gc}}  
\tablehead{  
  \colhead{Star}  
& \colhead{x\tablenotemark{a}}  
& \colhead{y\tablenotemark{a}}  
& \colhead{z$_{1}$\tablenotemark{a}}  
& \colhead{z$_{2}$\tablenotemark{a}}  
& \colhead{v}  
& \colhead{M$_\odot$} \\  
  
& \colhead{(arcsec)}  
& \colhead{(arcsec)}  
& \colhead{(arcsec)}  
& \colhead{(arcsec)}  
& \colhead{(km s$^{-1}$)}  
& \colhead{(${10}^{-5}\;M_\odot$ yr$^{-1}$)}  
}  
\startdata  
  
IRS 16NE                   &   \phn$-$2.6 &  \phn\phs0.8 & \phn\phs2.2 &  \phn\phs6.8 & \phn\phm{,}550 & \phn9.5 \\  
IRS 16NW                   &  \phn\phs0.2 &  \phn\phs1.0 &  \phn$-$8.3 &   \phn$-$5.5 & \phn\phm{,}750 & \phn5.3 \\  
IRS 16C                    &   \phn$-$1.0 &  \phn\phs0.2 & \phn\phs4.5 &  \phn\phs2.1 & \phn\phm{,}650 &    10.5 \\  
IRS 16SW                   &   \phn$-$0.6 &   \phn$-$1.3 &  \phn$-$2.5 &   \phn$-$1.2 & \phn\phm{,}650 &    15.5 \\  
IRS 13E1                   &  \phn\phs3.4 &   \phn$-$1.7 &  \phn$-$0.3 &  \phn\phs1.3 &          1,000 &    79.1 \\  
IRS 7W                     &  \phn\phs4.1 &  \phn\phs4.8 &  \phn$-$5.5 &   \phn$-$2.8 &          1,000 &    20.7 \\  
AF                         &  \phn\phs7.3 &   \phn$-$6.7 & \phn\phs6.2 &   \phn$-$1.2 & \phn\phm{,}700 & \phn8.7 \\  
IRS 15SW                   &  \phn\phs1.5 &     \phs10.1 & \phn\phs8.7 &  \phn\phs0.3 & \phn\phm{,}700 &    16.5 \\  
IRS 15NE                   &   \phn$-$1.6 &     \phs11.4 & \phn\phs0.7 &   \phn$-$1.1 & \phn\phm{,}750 &    18.0 \\  
IRS 29N\tablenotemark{b}   &  \phn\phs1.6 &  \phn\phs1.4 & \phn\phs8.3 &  \phn\phs3.2 & \phn\phm{,}750 & \phn7.3 \\  
IRS 33E\tablenotemark{b}   &  \phn\phs0.0 &   \phn$-$3.0 & \phn\phs0.6 &  \phn\phs6.0 & \phn\phm{,}750 & \phn7.3 \\  
IRS 34W\tablenotemark{b}   &  \phn\phs3.9 &  \phn\phs1.6 & \phn\phs4.0 &   \phn$-$4.8 & \phn\phm{,}750 & \phn7.3 \\  
IRS 1W\tablenotemark{b}    &   \phn$-$5.3 &  \phn\phs0.3 &  \phn$-$0.2 &   \phn$-$4.5 & \phn\phm{,}750 & \phn7.3 \\  
IRS 9NW\tablenotemark{b}   &   \phn$-$2.5 &   \phn$-$6.2 &  \phn$-$3.5 &   \phn$-$4.1 & \phn\phm{,}750 & \phn7.3 \\  
IRS 6W\tablenotemark{b}    &  \phn\phs8.1 &  \phn\phs1.6 & \phn\phs3.1 &   \phn$-$0.4 & \phn\phm{,}750 & \phn7.3 \\  
AF NW\tablenotemark{b}     &  \phn\phs8.3 &   \phn$-$3.1 &  \phn$-$0.1 &   \phn$-$2.4 & \phn\phm{,}750 & \phn7.3 \\  
BLUM\tablenotemark{b}      &  \phn\phs9.2 &   \phn$-$5.0 &  \phn$-$4.1 &  \phn\phs0.2 & \phn\phm{,}750 & \phn7.3 \\  
IRS 9S\tablenotemark{b}    &   \phn$-$5.5 &   \phn$-$9.2 &  \phn$-$5.9 &   \phn$-$0.3 & \phn\phm{,}750 & \phn7.3 \\  
Unnamed 1\tablenotemark{b} &  \phn\phs1.3 &   \phn$-$0.6 &  \phn$-$5.4 &  \phn\phs5.5 & \phn\phm{,}750 & \phn7.3 \\  
IRS 16SE\tablenotemark{b}  &   \phn$-$1.4 &   \phn$-$1.4 &  \phn$-$8.1 &   \phn$-$5.7 & \phn\phm{,}750 & \phn7.3 \\  
IRS 29NE\tablenotemark{b}  &  \phn\phs1.1 &  \phn\phs1.8 & \phn\phs3.1 &   \phn$-$3.1 & \phn\phm{,}750 & \phn7.3 \\  
IRS 7SE\tablenotemark{b}   &   \phn$-$2.7 &  \phn\phs3.0 & \phn\phs2.3 &   \phn$-$5.4 & \phn\phm{,}750 & \phn7.3 \\  
Unnamed 2\tablenotemark{b} &  \phn\phs3.8 &   \phn$-$4.2 &  \phn$-$8.5 &  \phn\phs4.5 & \phn\phm{,}750 & \phn7.3 \\  
IRS 7E\tablenotemark{b}    &   \phn$-$4.2 &  \phn\phs4.9 & \phn\phs8.6 &  \phn\phs1.3 & \phn\phm{,}750 & \phn7.3 \\  
AF NWW\tablenotemark{b}    &     \phs10.2 &   \phn$-$2.7 &  \phn$-$1.9 &  \phn\phs3.9 & \phn\phm{,}750 & \phn7.3 \\  
  
\enddata  
\tablenotetext{a}{Relative to Sgr A* in l-b coordinates where $+$x is  
  west and $+$y is north of Sgr A*}  
\tablenotetext{b}{Wind velocity and mass loss rate fixed (see text)}  
\end{deluxetable}  
  
\clearpage  
  
\clearpage  
  
\clearpage  
  
\clearpage  
  
\clearpage

\clearpage

\clearpage

\clearpage  
  
\end{document}